\documentclass[12pt]{article}
\usepackage{tabls}
\usepackage{amsmath,amssymb,graphicx,multirow,subfigure,oldgerm,euscript,mathrsfs}
\usepackage[section]{placeins}
\newcommand{\sdot}{\!\cdot\!}

\begin{document}

\begin{titlepage}
 
\begin{flushright}
\begin{tabular}{l}
  PCCF-RI-03-03 \\
  ADP-03-111/T549 \\
\end{tabular}  
\end{flushright}
 
\null\vskip 0.5 true cm
%
%
\begin{center}
{{\Large \bf Direct $\boldsymbol{CP}$ Violation In $\boldsymbol{B\rightarrow \pi^{+}\pi^{-} V}$  With
    $\boldsymbol{\rho^{0}-\omega}$ Mixing Effects: Phenomenological Approach }} \\
\vskip 1.0 cm
{ \Large Z.J. Ajaltouni$^1$\footnote{ziad@clermont.in2p3.fr},
O. Leitner$^{1,2}$\footnote{oleitner@physics.adelaide.edu.au},
P. Perret$^1$\footnote{perret@clermont.in2p3.fr},
C. Rimbault$^1$\footnote{rimbault@clermont.in2p3.fr},\\
A.W. Thomas$^2$\footnote{athomas@physics.adelaide.edu.au}} \\

\bigskip
{{\small \it $^1$ Laboratoire de Physique Corpusculaire de Clermont-Ferrand \\
IN2P3/CNRS Universit\'e Blaise Pascal \\
F-63177 Aubi\`ere Cedex France  \\
 $^2$ Department of Physics and Mathematical Physics and \\
 Special Research Centre for the Subatomic   Structure of Matter, \\ 
  University of Adelaide, 
  Adelaide 5005, Australia}}
\vskip 1 true cm
\end{center}
\vspace{1.5cm}


\begin{abstract}
We present a detailed study of direct $CP$ violation and
branching ratios in the channels $B^{0,\pm} \rightarrow 
\pi^{+}\pi^{-} V^{0,\pm}$, where $V$ is a vector meson ($K^{*
  0,\pm}$ or $\rho^{\pm}$). Emphasis is placed upon 
the important role played by ${{\rho}^{0}}-{\omega}$ mixing effects in the
estimation of the $CP$-violating asymmetry parameter, $a_{cp}$,
associated with the difference of $B$ and $\bar B$ decay amplitudes.
A thorough study of the helicity amplitudes is presented as a function of
the pion-pion invariant mass. All of the calculations and
simulations considered correspond to channels which will be analyzed 
at the LHCb facility.
\end{abstract}

\vspace{9.em}
\end{titlepage}

\newpage

%
\section{Introduction}\label{sec1}
%
Understanding the physical origin of the violation of $CP$ (Charge Conjugation
$\times$Parity) symmetry 
is one of the main goals of Particle Physics at the present time.
Recent experiments at $e^+ e^-$ colliders (BaBar, Belle) 
have produced fundamental 
results which strengthen the CKM picture of $CP$ violation~\cite{ref1,ref2} 
in the $B$ meson sector~\cite{Belle,Babar}.
However, the main results of these two collaborations are related to 
$B$ decays into pairs of pseudo-scalar mesons  
or into a vector plus a pseudo-scalar meson. 

A very broad physics program can also be carried out in the 
sector with two vector mesons in the final state, following 
$B$ decay. Apart from measuring the {\it standard angles},
$\alpha, \beta$ and $\gamma$ of the 
Unitary Triangle (UT), the vector mesons are {\it polarized} and 
their decay products (usually 
long-lived $0^{-+}$ mesons) are {\it correlated}. 
This opens the possibility of making interesting cross-checks of 
the Standard Model predictions as well testing some specific 
models of $CP$ violation beyond the SM approach (BSM). 

In the special case of two neutral vector mesons, the orbital angular   
momentum, $\ell$, and the total spin, $S$,  
of the $V^0_1 V^0_2$ system satisfy the equality $\ell = S = 0, 1, 2$. 
The $CP$ eigenvalues  
are defined as ${(-1)}^{\ell}$. Because of the allowed values for the angular
momentum, $\ell$, one has a very clear indication of any
{\it mixing} of different $CP$
eigenstates and hence of $CP$ non-conservation. The
separation of the different $CP$ eigenstates requires a detailed analysis of the
final angular distributions~\cite{Dunietzetal}. However, because this
analysis can be carried out in
a model-independent way, it provides a significant constraint on
any model.

After explaining the helicity formalism (Section II), 
a special study is devoted to the final state interactions
(FSI) and the key role of ${{\rho}^0} - {\omega}$ mixing (Section III). A
complete and realistic
determination of the helicity amplitudes, in the framework of the effective
Hamiltonian approach, is introduced in Section IV.
Then, the main results of the Monte-Carlo simulations,
providing estimates of the various density
matrix elements
$h_{ij}$, are shown in Section V. In the following section (Section VI)
the numerical analysis and discussions about the branching 
ratios and asymmetries for
$B$ decays into two vectors ($B \rightarrow \rho^0(\omega) V_2$, with $V_2=
K^{* 0}, {\bar K^{* 0}}, K^{* -}, K^{* +}, \rho^+, \rho^- $)
are given in detail. 
These two vectors, $\rho^0(\omega)$ and $V_2$, each decay into two
pseudo-scalars. Emphasis is put on the angular distributions of the
pseudo-scalar mesons in both the helicity and 
transversity frames. Finally, 
in the last section, we summarize our results for the
different channels which will be investigated in future experiments at   
$p {\bar p}$ colliders and make some concluding remarks. 
%
\section{General formalism for $\boldsymbol{B \rightarrow V_1 V_2}$ decays}\label{sec2}
%
%
\subsection{Helicity frame}\label{sec2.1}
%
Because the $B$ meson has spin $0$, the final two vector 
mesons, $V_1$ and $ V_2$, have the same helicity, 
${\lambda}_1 = {\lambda}_2 = -1,0, +1,$ and their 
angular distribution is isotropic in the $B$ rest frame.
Let $H_w$ be the weak Hamiltonian which governs the $B$ decays. Any transition
amplitude between the initial and final states will have the following form:
\begin{equation}\label{eq1}
 H_{\lambda} = \langle V_1{(\lambda)} V_2{(\lambda)}|H_w|B \rangle\ ,
\end{equation} 
where the common helicity is ${\lambda} = -1,0, +1$.  
Then, each vector meson $V_i$ will decay into two 
pseudo-scalar mesons, $a_i,b_i$,
where  $a_i$ and $b_i$ can be either a pion or a kaon, and the angular
distributions of $a_i$ and $b_i$  depend on the polarization of $V_i$.

The helicity frame of a vector-meson $V_i$ is defined in the $B$ 
rest frame  
such that the direction of the Z-axis is given by its momentum, 
${\vec{p}_i}$. Schematically, the whole process  gets the form:
\begin{equation*}
 B     \longrightarrow     V_1  +   V_2  \longrightarrow   (a_1 + b_1)  +
 (a_2 + b_2)\ .
\end{equation*}      
The corresponding decay amplitude, $M_{\lambda}\bigl(B \rightarrow \sum_{i=1}^2
(a_i+b_i)\bigr)$, is factorized out according to the relation,
\begin{equation}\label{eq2}
 M_{\lambda}\bigl(B \rightarrow \sum_{i=1}^2 (a_i+b_i)\bigr) = 
  H_{\lambda}(B \rightarrow V_1 +V_2) \times \prod_{i=1}^2 A_i(V_i \rightarrow a_i + b_i)\ ,    
\end{equation}
where the amplitudes $A_i(V_i \rightarrow a_i + b_i)$ are 
related to the decay of the resonances
$V_i$. The $A_i(V_i \rightarrow a_i + b_i)$  are  
given by the following expressions:
\begin{align}\label{eq3}
A_1(V_1 \rightarrow a_1 + b_1) & =   \sum_{m_1=-1}^1  c_1 
D^1_{\lambda,m_1}(0,\theta_1,0)\ ,  \nonumber  \\
A_2(V_2 \rightarrow a_2 + b_2) &  =   \sum_{m_2=-1}^1  c_2 
D^1_{\lambda,m_2}(\phi,\theta_2,-\phi)\ .  
\end{align} 
These equalities are an illustration of the Wigner-Eckart theorem. In
Eq.~(\ref{eq3}), the $c_1$
and $c_2$ parameters represent, respectively,
the {\it dynamical  decays} of the $V_1$
and $V_2$ resonances. 
The term $D^1_{\lambda,m_i}(\phi_i,\theta_i,-\phi_i)$ is the 
Wigner rotation matrix element for a spin-1 particle and we
let $\lambda{(a_i)}$ and $\lambda{(b_i)}$ be the  
respective helicities of the final particles $a_i$ and $b_i$
in the $V_i$ rest frame. $\theta_1$  is the polar angle of
$a_1$ in the $V_1$ helicity frame. 
The decay plane of $V_1$ is identified with  
the (X-Z) plane and consequently the azimuthal angle 
$\phi_1$ is set to $0$. Similarly,
$\theta_2$ and $\phi$ are respectively the polar and azimuthal angles of
particle $a_2$ in the $V_2$ helicity frame. Finally, the coefficients $m_i$
are defined as:
\begin{equation}\label{eq4}
m_i=\lambda(a_i)-\lambda(b_i)\ .
\end{equation} 
Our convention for the $D^1_{\lambda,m_i}(\alpha,\beta,\gamma)$ 
matrix element 
is given in Rose's book~\cite{Rose}, namely: 
\begin{equation}\label{eq5}
D^1_{\lambda,m_i}(\alpha,\beta,\gamma) =  \exp[-i(\lambda \alpha +
m_i \gamma)] \;  d^1_{\lambda,m_i} (\beta)\ .
\end{equation}
The most general form of the decay amplitude ${\cal M}\bigl(B \rightarrow \sum_{i=1}^2
(a_i + b_i)\bigr)$ is a {\it linear superposition} 
of the previous amplitudes $M_{\lambda}\bigl(B \rightarrow \sum_{i=1}^2 (a_i
+b_i)\bigr)$ denoted by,
\begin{equation}\label{eq6}
 {\cal M}\bigl(B \rightarrow \sum_{i=1}^2 (a_i +b_i)\bigr)= 
  \sum_{\lambda} M_{\lambda}\bigl(B \rightarrow \sum_{i=1}^2 (a_i +b_i)\bigr)\ . 
\end{equation} 
The decay width, $\Gamma{(B \rightarrow V_1 V_2)}$, can be
computed by taking the square of the modulus,  
$\Bigl|{\cal M}\bigl(B \rightarrow \sum_{i=1}^2 (a_i +b_i)\bigr)\Bigr|$,
which involves the three kinematic parameters 
$\theta_1, \theta_2$ and  $\phi$. This leads to the following
general expression: 
\begin{equation}\label{eq7}
d^3{\Gamma}(B \rightarrow V_1 V_2)  \propto  \Bigl|\sum_{\lambda}
M_{\lambda} \bigl( B \rightarrow
\sum_{i=1}^2 (a_i +b_i)\bigr)\Bigr|^2  =  \sum_{\lambda,\lambda'}
 h_{\lambda, \lambda'} F_{\lambda,\lambda'}(\theta_1) G_{\lambda,\lambda'}
 (\theta_2, \phi)\ ,
\end{equation} 
which involves three density-matrices,
$h_{\lambda, \lambda'}, F_{\lambda,
\lambda'}(\theta_1)$ and  $G_{\lambda, \lambda'}(\theta_2, \phi)$.
The factor $h_{\lambda, \lambda'}=H_{\lambda} H^{*}_{\lambda'}$ is an 
element of the density-matrix related to the $B$ decay, while $F_{\lambda,
\lambda'}(\theta_1)$ represents the density-matrix of 
the decay $V_1 \rightarrow a_1 + b_1$.
In a similar way, $G_{\lambda, \lambda'}(\theta_2, \phi)$ represents the
density-matrix of the decay $V_2 \rightarrow a_2+b_2$.
 
The analytic expression in Eq.~(\ref{eq7}) exhibits 
a very general form. It depends on neither 
the specific nature of the intermediate resonances nor their 
decay modes (except for the spin of the final particles). 
This approach also presents three key advantages.
The first one comes from the fact that all the dynamics of the $B$ decay is
introduced into the coefficients 
$h_{{\lambda},{\lambda}'}$. This allows us to use various theoretical models
involving different dynamical processes and form factors. The second is
associated with the fact that the formal 
expressions for $F_{\lambda, \lambda'}(\theta_1)$
and $G_{\lambda, \lambda'}(\theta_2, \phi)$, which are related to the
polarization of the intermediate resonances, remain unchanged whatever
the coefficients $h_{\lambda, \lambda'}$ happen to be.
Finally, correlations among final particles arise in a straightforward way
because of the previous expression which relates the angles ${\theta}_1,
{\theta}_2$ and $\phi$. Consequently, a probability 
density function (pdf) can be inferred from the
general expression and one gets:
\begin{equation}\label{eq8} 
  f(\theta_1 , \theta_2 , \phi)  =  \frac {d^3\Gamma(B \rightarrow V_1 V_2)}{\Gamma(B \rightarrow V_1 V_2)
  d{(\cos\theta_1)} d{(\cos\theta_2)} d\phi}\ ,    
\end{equation}
where the angles $\theta_1,\theta_2$ and $\phi$   
were defined earlier and 
$\Gamma(B \rightarrow V_1 V_2)$ is the partial decay width. 
This function allows one to compute
three other pdfs separately for the variables $\theta_1, \theta_2$ and $\phi$.

The previous calculations are illustrated by the reaction $B^0 \rightarrow
K^{* 0} {\rho}^0$ where $K^{* 0} \rightarrow K^+ {\pi}^-$ and 
${\rho}^0 \rightarrow {\pi}^+ {\pi}^{-}$. In this channel, since all the final
particles have spin zero,  the coefficients  $m_1$ and $m_2$, defined
in Eq.~(\ref{eq4}), are equal to zero. The three-fold differential
width has the following form:
\begin{multline}\label{eq9}
\frac{d^3\Gamma(B \rightarrow V_1 V_2)}{d(\cos\theta_1) d(\cos\theta_2) d\phi}  \propto 
\bigl(h_{++} + h_{--}\bigr){{\sin}^2{\theta_1}{\sin}^2{\theta_2}}/4 +
{h_{00}{\cos}^2{\theta_1}{\cos}^2{\theta_2}} \\
+ \Bigl\{\mathscr{R}\!e{(h_{+0})}{\cos{\phi}} - \mathscr{I}\!m{(h_{+0})}{\sin{\phi}} + \mathscr{R}\!e{(h_{0-})}{\cos{\phi}} -
\mathscr{I}\!m{(h_{0-})}{\sin{\phi}}\Bigr\}{{\sin{2\theta_1}}{\sin{2\theta_2}}}/4  \\
+ \Bigl\{\mathscr{R}\!e{(h_{+-})}{\cos{2\phi}} - \mathscr{I}\!m{(h_{+-})}{\sin{2\phi}}\Bigr\}
{{\sin}^2{\theta_1}{\sin}^2{\theta_2}}/2\ ,
\end{multline}
where all the terms in Eq.~(\ref{eq9}) have been already specified.
It is worth noticing that the  expression in Eq.~(\ref{eq9}) is {\it
completely symmetric} in ${\theta}_1$ and ${\theta}_2$ and consequently, the
angular distributions of $a_1$ in the $V_1$ frame is identical that of
$a_2$ in the $V_2$ frame. From Eqs.~(\ref{eq8}) and~(\ref{eq9})  
the normalized pdfs of ${\theta}_1$, 
${\theta}_2$ and $\phi$ can be derived and one finds:
\begin{align}\label{eq10}  
 f{({\cos}{\theta}_{1,2})} & =  {(3h_{00}-1)}{{\cos}^2{\theta}_{1,2}}  +  {(1-
 h_{00})}\ ,  \nonumber \\
 g{(\phi)} & =   1 + 2 \; \mathscr{R}\!e{(h_{+-})}{\cos{2\phi}}  -  2 \; \mathscr{I}\!m{(h_{+-})}{\sin{2\phi}}\ . 
\end{align}
%
%
\subsection{Transversity frame}\label{sec2.2}
\begin{figure}
\centering\includegraphics[height=20.0em, width = 34.0em,clip=true]{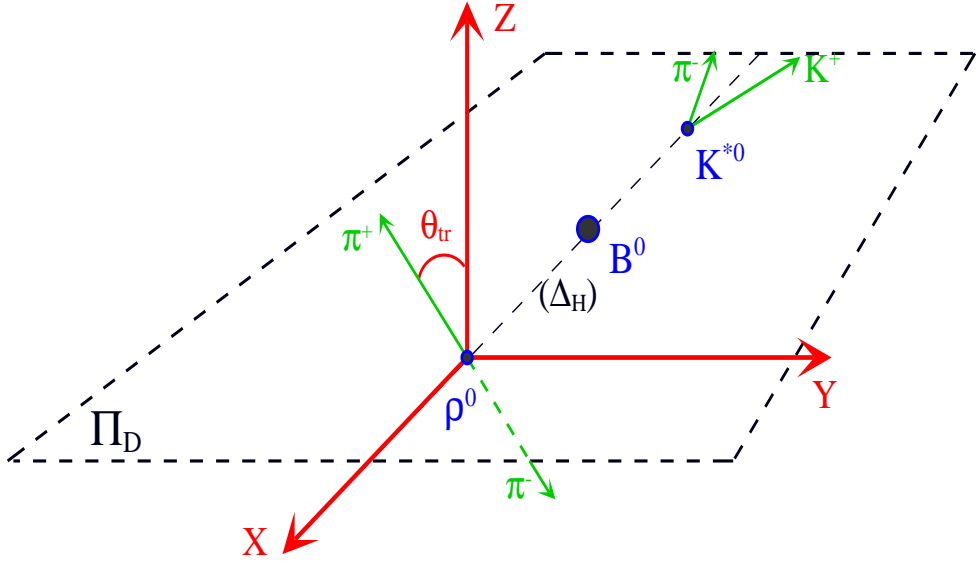}
\caption{Transversity frame for $B \to \rho^{0} K^{*}$.}
\label{figOO1}
\end{figure}

%
Initially, the transversity frame (TF) was introduced by
A. Bohr~\cite{Abohr} in order to facilitate the determination of the 
spin and parity of a resonance decaying into 
stable particles. It can be extended
to a system of two vector mesons coming from a heavy meson, $B$ or $\bar B$,
in order to perform tests of $CP$ symmetry. 
In displaying new angular distributions, the TF provides 
complementary physical information to that seen in   
the standard helicity frame. The construction of 
the TF and its use require several steps. For a clear
illustration,  see Fig.~\ref{figOO1}, where the channel $B^0 \rightarrow
{\rho}^0 K^{* 0}$  is chosen.

Departing from the $B$ rest frame, 
the common helicity axis, $({\Delta}_H)$, is given by the direction of the
momentum ${\vec {p_1}}$. This and the decay 
plane, $({\Pi}_D)$, of the vector meson 
($K^{* 0} \rightarrow {K^+} {{\pi}^-}$) 
are the main ingredients of the TF.
The vector meson ${\rho}^0$ is taken at rest 
(origin of the frame) and the X-axis 
is given by  $({\Delta}_H)$. In the decay 
plane, $({\Pi}_D)$, the Y-axis which
is orthogonal to the X-axis, is chosen in such a way
that the ${K}^+$ meson has the Y-component of its momentum 
greater than or equal to zero.
The Z-axis, which is orthogonal to the plane $({\Pi}_D)$,
is obtained by the classical relation
${\vec e}_Z = {{\vec e}_X} {\times} {{\vec e}_Y}$.

The angular distributions of the  ${\pi}^{\pm}$ coming from 
the $\rho^{0}$
decay are referred to the new Z-axis. 
It is worthy noticing that, in the TF, the flying meson 
and its decay products are very energetic compared to the $B$ frame. 
Explicitly, the $\rho^{0}$ energy is given by the relation,
\begin{equation}\label{eq11}
E_{\rho^{0}}  =  {({m_B}^2 - {m_1}^2 - {m_2}^2)}/{2{m_1}}   \approx    17  \; {\rm GeV}\ , 
\end{equation}   
where $m_1$ and $m_2$ refer to the 
masses of the   $K^{* 0}$ and ${\rho}^0$ resonances, respectively.
As far as the transition amplitudes in the TF are concerned, 
they are a simple linear combination
of the helicity amplitudes, namely:
\begin{equation}\label{eq12}
 H_P  =  \frac{H_+ + H_-}{\sqrt 2}\ , \; {\rm and} \;\; H_T = \frac{H_+ - H_-}{\sqrt 2}\ , 
\end{equation} 
while $H_0$ remains unchanged. We can rewrite the angular distributions given
in Eq.~(\ref{eq10}) by using the relations from  Eq.~(\ref{eq12}) and angles
$\theta_{1,2}, \phi$ expressed in the transversity frame. Thus one gets,
\begin{align}\label{eq13}
 f_T{({\cos}{\theta_{1,2}})}  & =  {\bigl(3{|H_T|}^2-1\bigr)}{{\cos}^2{\theta_{1,2}}}  + 
 \bigl(1-{|H_T|}^2\bigr)\ , \nonumber \\
 g_T{(\phi)} & =  {\bigl(1 + {|H_0|}^2 -{|H_P|}^2\bigr)} {\cos{2\phi}}\ .  
\end{align} 
%
\section{Final state interactions and $\boldsymbol{{{\rho}^0}-{\omega}}$ mixing}\label{sec3}
%
\subsection{Factorization hypothesis}\label{sec3.1}
%
{}Final state interactions (FSI) represent unavoidable 
effects in hadronic physics and they play a crucial
role in heavy resonance decays \cite {Hquinn}. In the case of a $B$ meson, 
characterized by a center-of-mass energy 
$\sqrt s \approx    5.3$ GeV, the charmless weak decays of the b-quark lead
to {\it light energetic} quarks
 which can exchange several gluons amongst
themselves as well as with the spectator quark in the $B$ meson. 
This fundamental process occurs in decays described 
by tree, penguin and annihilation diagrams 
and is characterized by two regimes: 
perturbative and non-perturbative. 
In order to handle the FSI in both regimes, 
the usual method is inspired by the effective Lagrangian
approach. Perturbative calculations at
next-to-leading order (NLO) are performed for a scale higher than $m_b$ (since
our analysis is focused on $B$ decays) and the 
non-perturbative effects are inserted for a scale lower than $m_b$. 
This general method is called the {\it factorization procedure}~\cite{ref19}  
and further details are given below. 

In the factorization approximation, either the vector meson $\rho^{0}(\omega)$ or the
$K^{*}$ is generated by one current in the effective Hamiltonian 
which has the appropriate quantum numbers.
For the $B$ decay processes considered here, two kinds of 
matrix element products are involved after factorization 
(i.e.~omitting Dirac matrices and color labels):  
$\langle \rho^{0}(\omega)|(\bar{u}u)|0\rangle 
\langle K^{*}|(\bar{q}_{i}q_{j})|B^{\pm,0}\rangle $ and $ \langle K^{*}|
(\bar{q}_{i}q_{j})|0\rangle \langle\rho^{0}
(\omega)|(\bar{u}b)|B^{\pm,0}\rangle$, where $q_{i}$ and $q_{j}$ could be 
either $u,s$ or $d$. We will 
calculate them in two phenomenological quark models. 

The matrix elements for  $B \rightarrow X^{*}$ (where  $ X^{*}$ 
denotes a vector meson) can be decomposed as follows~\cite{ref20},
\begin{multline}\label{eq14}
\langle X^{*}|J_{\mu}|B \rangle=\frac{2}{m_{B}+m_{X^{*}}} \epsilon_{\mu \nu \rho \sigma} 
\epsilon^{* \nu} p_{B}^{\rho} p_{X^{*}}^{\sigma}  V(k^{2}) 
+i \Biggl\{ \epsilon_{\mu}^{*}(m_{B}
+m_{X^{*}})A_{1}(k^{2})  \\
- \frac{\epsilon^{*}  \cdot k}{m_{B}+m_{X^{*}}} (P_{B}+P_{X^{*}})_{\mu}A_{2}(k^{2})- \frac 
{\epsilon^{*}  \cdot k}{k^{2}}2m_{X^{*}} \cdot k_{\mu}A_{3}(k^{2})
\Biggr\}  \\
+i \frac{\epsilon^{*} \cdot k}{ k^{2}}2m_{X^{*}} \cdot k_{\mu}A_{0}(k^{2})\ ,
\end{multline}
where $J_{\mu}$  is the weak current, defined as 
$J_{\mu}=\bar{q}\gamma^{\mu}(1-\gamma_{5})b \;\;     
{\rm with} \;\; q=u,d,s$ and  $k=p_{B}-p_{X^{*}}$ and  
$\epsilon_{\mu}$ is the polarization 
vector of $X^{*}$. The form factors $A_{0}, A_{1}, A_{2}, A_{3}$ 
and $ V$ describe the 
transition $0^{-} \rightarrow 1^{-}$.
{}Finally, in order to cancel the poles at $q^{2}=0$, 
the form factors respect the condition:
\begin{equation}\label{eq15}
 A_{3}(0) = A_{0}(0)\ ,
\end{equation}
and they also  satisfy the following relations: 
\begin{gather}\label{eq16}
A_{3}(k^{2})  = \frac {m_{B}+m_{X^{*}}}{2m_{X^{*}}}A_{1}(k^{2})- \frac {m_{B}-m_{X^{*}}}
{2m_{X^{*}}}A_{2}(k^{2})\ . 
\end{gather}
In the evaluation of matrix elements, the effective number of colors,
 $N_{c}^{eff}$, enters through  a Fierz 
transformation. In general, for an operator $O_{i}$, one can write,
\begin{equation}\label{eq17}
\frac{1}{(N_{c}^{eff})_{i}} = \frac{1}{3} + \xi_{i}\ , {\rm with} \; i=1, \cdots ,10 \ ,
\end{equation}
here $\xi_{i}$ describes non-factorizable effects.
$\xi_{i}$ is assumed to be universal for all the  
operators $O_{i}$. Naive factorization assumes that we can replace -in a heavy quark decay-
the matrix element of a four fermion operator by the product of the matrix elements
of two currents. This reduces to the product of a form factor and a decay
constant. This assumption is only rigorously justified at large values of $N_{c}$. But
it is known that naive factorization may give a good estimate of the magnitude of
the $B$ decay amplitude in many cases~\cite{refBBNS}.
%
\subsection{FSI at the quark level: strong phase generated by the penguin diagrams}\label{sec3.2}
%
Let $A$ be the amplitude for the decay 
$B \rightarrow \rho^{0} ( \omega ) K^{*} \rightarrow  \pi^{+} 
\pi^{-} K^{*}$ (a similar procedure applies in the case where we
have a $\rho^{\pm}$ \cite{Gardner:1997yx}
instead of the $K^{*}$), then one has,
\begin{equation}\label{eq18}
A=\langle  K^{*}  \pi^{-} \pi^{+}|H^{T}|B \rangle + \langle  K^{*}  
\pi^{-} \pi^{+}|H^{P}|B  \rangle\ ,
\end{equation}
with $H^{T}$ and $H^{P}$ being the Hamiltonians for 
the tree and penguin operators. We can 
define the relative magnitude and phases between these 
two contributions as follows,
\begin{align}\label{eq19}
A &= \langle  K^{*}  \pi^{-} \pi^{+}|H^{T}| B \rangle [ 1+re^{i\delta}e^{i\phi}]\ , \nonumber \\   
\bar {A} &= \langle \bar{K^{*}}  \pi^{+}  \pi^{-}|H^{T}|\bar {B} 
\rangle [ 1+re^{i\delta}e^{-i\phi}]\ ,  
\end{align}
where $\delta$ and $\phi$ are strong and weak phases, respectively. 
The phase $\phi$ arises from the 
appropriate combination of CKM matrix elements,    
$\phi={\rm arg}[(V_{tb}V_{ts}^{*})/(V_{ub}V_{us}^{*})]$. 
As a result, $\sin \phi$ is equal 
to $\sin \gamma$, with $\gamma$ defined in the standard 
way~\cite{ref16}. The parameter, $r$, is the 
absolute value of the ratio of tree and penguin amplitudes:
\begin{equation}\label{eq20}
r \equiv \left| \frac{\langle \rho^{0}(\omega) K^{*}|H^{P}|B \rangle}{\langle\rho^{0}(\omega)
K^{*}|H^{T}|B \rangle} \right|.
\end{equation}
%
%
\subsection{Strong phase generated by the $\boldsymbol{\rho^{0}-\omega}$ mixing}\label{sec3.3}
%
In the vector meson dominance model~\cite{refa1}, the photon propagator is
dressed by coupling to  vector mesons.
{}From this, the $\rho^{0}-\omega$ mixing 
mechanism~\cite{refa2}  was developed. 
In order to obtain a large signal for direct $CP$ 
violation, we need some mechanism to make both 
$\sin\delta$ and $r$ large. We stress that $\rho^{0}-\omega$ mixing has the
dual advantages that the strong 
phase difference is large (passing through $90^{o}$ at the $\omega$ resonance)
and well known~\cite{Gardner:1997yx,ref5}.
With this mechanism, to first order in isospin violation, 
we have the following results 
when the invariant mass of the $\pi^{+}\pi^{-}$ pair 
is near the mass of the $\omega$ resonance,
\begin{align}\label{eq21}
\langle K^{*} \pi^{-} \pi^{+}|H^{T}|B  \rangle & = \frac{g_{\rho}}{s_{\rho}s_{\omega}}
 \tilde{\Pi}_{\rho \omega}t_{\omega} +\frac{g_{\rho}}{s_{\rho}}t_{\rho}\ , \nonumber \\
\langle  K^{*} \pi^{-} \pi^{+}|H^{P}|B  \rangle & = \frac{g_{\rho}}{s_{\rho}s_{\omega}} 
\tilde{\Pi}_{\rho \omega}p_{\omega} +\frac{g_{\rho}}{s_{\rho}}p_{\rho}\ .
\end{align}
Here $t_{V} \; (V=\rho \;{\rm  or} \; \omega) $ is the tree amplitude and
$p_{V}$ the penguin amplitude for 
producing a vector meson, $V$, $g_{\rho}$ is the 
coupling for $\rho^{0} \rightarrow \pi^{+}\pi^{-}$,
$\tilde{\Pi}_{\rho \omega}$ is the effective $\rho^{0}-\omega$ mixing
amplitude, and $s_{V}$ is the inverse 
propagator of the vector meson $V$,
\begin{equation}\label{eq22}
s_{V}=s-m_{V}^{2}+im_{V}\Gamma_{V}\ , 
\end{equation}
with $\sqrt s$ being the invariant mass of the $\pi^{+}\pi^{-}$ pair. 
We stress that the direct coupling 
$\omega \rightarrow \pi^{+} \pi^{-} $ is effectively absorbed into 
$\tilde{\Pi}_{\rho \omega}$~\cite{ref17}, leading to the explicit $s$
dependence of $\tilde{\Pi}_{\rho \omega}$. 
Making the expansion $\tilde{\Pi}_{\rho \omega}(s)=
\tilde{\Pi}_{\rho \omega}(m_{\omega}^{2})+(s-m_{w}^{2}) 
\tilde{\Pi}_{\rho \omega}^{\prime}(m_{\omega}^{2})$, the  $\rho^{0}-\omega$ mixing
parameters were determined in 
the fit of Gardner and O'Connell~\cite{ref18}: $\mathscr{R}\!e \; 
\tilde{\Pi}_{\rho \omega}(m_{\omega}^{2})=-3500 \pm 300 \; {\rm MeV}^{2},
\;\;\; \mathscr{I}\!m \; \tilde{\Pi}_{\rho \omega}
(m_{\omega}^{2})= -300 \pm 300 \; {\rm MeV}^{2}$, and  $\tilde{\Pi}_{\rho \omega}^{\prime}
(m_{\omega}^{2})=0.03 \pm 0.04$. In practice, the effect of the derivative term is negligible.
{}From Eqs.~(\ref{eq18}) and (\ref{eq21}) one has
\vspace{0.5em}
\begin{equation}\label{eq23}
 re^{i \delta} e^{i \phi}= \frac{ \tilde {\Pi}_{\rho \omega}p_{\omega}+s_{\omega}p_{\rho}}{\tilde 
{\Pi}_{\rho \omega} t_{\omega} + s_{\omega}t_{\rho}}\ . 
\end{equation}
Defining
\vspace{-1.5em}
\begin{center}
\begin{equation}\label{eq24}
\frac{p_{\omega}}{t_{\rho}} \equiv r^{\prime}e^{i(\delta_{q}+\phi)}\ , \;\;\;\;
\frac{t_{\omega}}{t_{\rho}} \equiv \alpha e^{i \delta_{\alpha}}\ , \;\;\;\;
\frac{p_{\rho}}{p_{\omega}} \equiv \beta e^{i \delta_{\beta}}\ , 
\end{equation}
\end{center}
where $ \delta_{\alpha}, \delta_{\beta}$ and $ \delta_{q}$ are partial strong phases (absorptive part) 
arising from the tree and penguin diagram contributions.
Substituting Eq.~(\ref{eq24}) into  Eq.~(\ref{eq23}), one finds:
\vspace{0.5em}
\begin{equation}\label{eq25}
re^{i\delta}=r^{\prime}e^{i\delta_{q}} \frac{\tilde{\Pi}_{\rho \omega}+ \beta e^{i \delta_{\beta}} 
s_{\omega}}{s_{\omega}+\tilde{\Pi}_{\rho \omega} \alpha e^{i \delta_{\alpha}}}\ ,
\end{equation}
where the total strong phase, $\delta$, is mainly proportional to the 
ratio of the penguin and tree diagram contributions.
%
\subsection{Importance of the strong phase for $\boldsymbol{B \bar{B}}$ asymmetry}\label{sec3.4}
%
Under a $CP$ transformation the strong phase, $\delta$, remains 
unchanged, while the weak phase, $\phi$, 
which is related to the CKM matrix elements, changes sign.
 Thus, the asymmetry parameter, $a_{CP}^{dir}$, 
which can reveal {\it direct $CP$ violation}, can be 
deduced in the following way:
\begin{eqnarray}\label{eq26}
a_{CP}^{dir} = {\frac {A^2 -{\bar A}^2}{A^2 +{\bar A}^2}} = \frac{-2 \ r \ {\sin {\delta}} \ {\sin {\phi}}}{1 + r^2
+2 \ r \ {\cos {\delta}} \ {\cos {\phi}}}\ . 
\end{eqnarray}

It is straightforward to see that the parameter $a_{CP}^{dir}$ 
depends on both the strong phase 
{\it and} the weak phase and, consequently, that 
the maximum value of $a_{CP}^{dir}$ can be reached if $\sin {\delta} = 1$. 
This is why
the strong final state interaction (FSI) among pions 
coming from ${\rho}^0 - \omega$ mixing
{\it enhances} the
direct $CP$ violation in the vicinity of the mass of the 
$\omega$ resonance.

In the Wolfenstein parametrization~\cite{ref15}, the weak phase comes from 
$[V_{tb} V^{*}_{ts}/V_{ub} V^{*}_{us}]$ and one has for the decay 
$B \rightarrow \rho^{0}(\omega) K^{*}$,
\begin{align}\label{eq27a}
\sin\phi & = \frac{-\eta}{\sqrt {\rho^2+\eta^{2}}}\ , \nonumber \\
\cos\phi & = \frac{-\rho}{\sqrt {\rho^2+\eta^{2}}}\ ,
\end{align}
while the weak phase comes from 
$[V_{tb} V^{*}_{td}/V_{ub} V^{*}_{ud}]$ for the decay 
$B \rightarrow \rho^{0}(\omega) \rho$,
\begin{align}\label{eq27b}
\sin\phi & = \frac{\eta}{\sqrt {[\rho(1-\rho)- \eta^{2}]^{2}+\eta^{2}}}\ ,
\nonumber \\
\cos\phi & = \frac{\rho(1-\rho)-\eta^{2}}{\sqrt {[\rho(1-\rho)-\eta^{2}]^{2}+\eta^{2}}}\ .
\end{align}
The values used for $\rho$ and $\eta$ will be discussed in Section V.
%
\section{Explicit calculations according to the effective \\ Hamiltonian}\label{sec4}
%
\subsection{Generalities concerning the OPE for weak hadronic decays}
\label{sec4.1}
%
\subsubsection{Operator product expansion }\label{sec4.1.1}
%
The operator product expansion (OPE)~\cite{ref6} is an extremely
useful tool in the analysis of weak interaction processes involving 
quarks. Defining the decay amplitude $A(M \rightarrow F)$ as
\begin{equation}\label{eq28}
A(M \rightarrow F) \propto C_{i}(\mu) \langle F | O_{i}(\mu) | M \rangle \ ,
\end{equation}
where $C_{i}(\mu)$ are the Wilson coefficients 
(see Section~\ref{sec4.1.2}), $O_{i}(\mu)$ are 
the operators given by the OPE and $\mu$ is an energy scale, one sees that
the OPE separates the calculation of the amplitude, $A(M \rightarrow F)$,  into
two distinct physical regimes. One is related to {\it hard} or 
short-distance physics, represented by $C_{i}(\mu)$ and calculated by a
perturbative approach. The other is 
the {\it soft} or long-distance regime. This part must be treated by   
non-perturbative approaches such as the $1/N$ expansion~\cite{ref7}, 
QCD sum rules~\cite{ref8},  
hadronic sum rules or lattice QCD.

\begin{figure}
\centering\includegraphics[height=17.0em, width = 34.0em,clip=true]{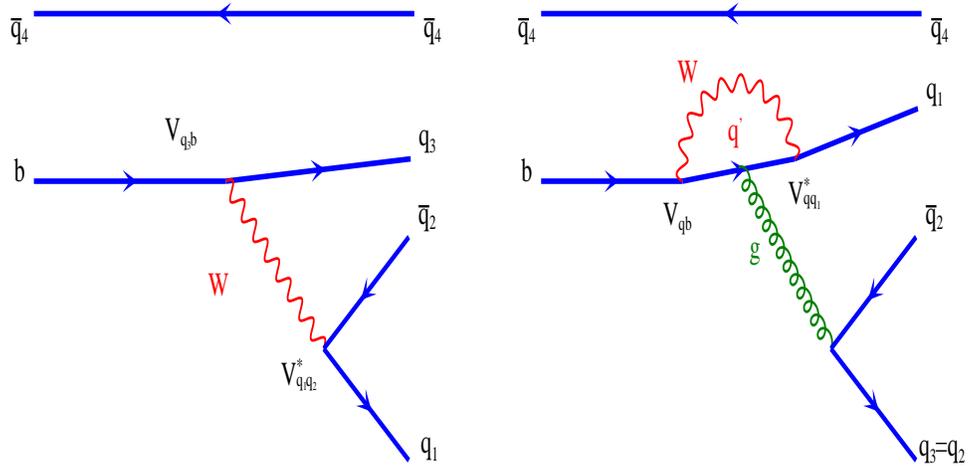}
\caption{Tree diagram (left), and  QCD-penguin diagram (right), for $B$ decays.}
\label{figO1}
\end{figure}

\begin{figure}
\centering\includegraphics[height=16.0em, width = 34.em,clip=true]{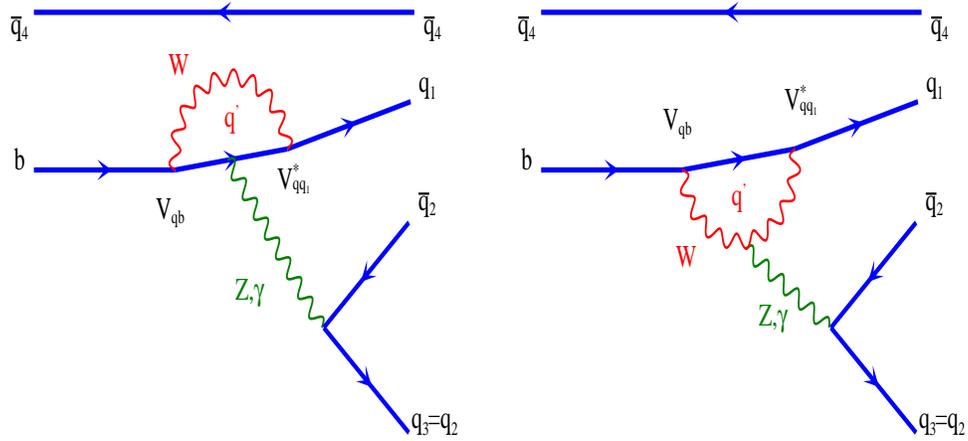}
\caption{Electroweak-penguin diagram
    (left), and electroweak-penguin diagram with coupling between $Z,\gamma$ and
    $W$ (right), for $B$ decays.}
\label{figO2}
\end{figure}

The operators, $O_{i}$, are local operators which 
can be written in the general form:
\begin{equation}\label{eq29}
O_{i} = ({\bar q}_{i} \Gamma_{n1} q_{j})({\bar q}_{k} \Gamma_{n2} q_{l})\ , 
\end{equation}
where $\Gamma_{n1}$ and $\Gamma_{n2}$ denote
a  combination of gamma matrices and $q$ the quark flavor. They  should respect
the Dirac structure, the 
color structure and the types of quarks relevant for the  decay being
 studied. They can be divided into two
classes according to topology:
tree  operators   ($O_{1}, O_{2}$), and  
penguin operators  ($O_{3}$ to $O_{10}$).
{}For tree contributions (where $W^{\pm}$ is exchanged), 
the Feynman diagram is shown
in Fig.~\ref{figO1} (left). The 
current-current operators related to the tree diagram are the following:
\begin{align}\label{eq30}
O_{1}^{s}& = \bar{q}_{\alpha}
\gamma_{\mu}(1-\gamma{_5})u_{\beta}\bar{s}_{\beta} \gamma^{\mu}(1-\gamma{_5})
b_{\alpha}\ , \nonumber \\
O_{2}^{s}& = \bar{q} \gamma_{\mu}(1-\gamma{_5})u\bar{s} \gamma^{\mu}(1-\gamma{_5})b\ , 
\end{align}
where $\alpha$ and $\beta$ are the color indices. 
The penguin terms can be divided 
into two sets. The first is from the QCD penguin 
diagrams where gluons are exchanged, while the
second is from  the 
electroweak penguin diagrams (where either a $\gamma$ or a $Z^{0}$ 
is exchanged). The Feynman
diagram for the QCD penguin diagram
is shown in
{}Fig.~\ref{figO1} (right) and the corresponding operators  
are written as follows:
\begin{align}\label{eq31a}
O_{3}& = \bar{q} \gamma_{\mu}(1-\gamma{_5})b \sum_{q\prime}\bar{q}^{\prime}\gamma^{\mu}(1-\gamma{_5})
q^{\prime}\ , \nonumber \\ 
O_{4}& =\bar{q}_{\alpha} \gamma_{\mu}(1-\gamma{_5})b_{\beta} 
\sum_{q\prime}\bar{q}^{\prime}_{\beta}\gamma^{\mu}(1-\gamma{_5})q^{\prime}_{\alpha}\ , 
\end{align}
\begin{align}\label{eq31b}
O_{5}& =\bar{q} \gamma_{\mu}(1-\gamma{_5})b \sum_{q'}\bar{q}^
{\prime}\gamma^{\mu}(1+\gamma{_5})q^{\prime}\ , \nonumber \\ 
O_{6}& =\bar{q}_{\alpha} \gamma_{\mu}(1-\gamma{_5})b_{\beta} 
\sum_{q'}\bar{q}^{\prime}_{\beta}\gamma^{\mu}(1+\gamma{_5})q^{\prime}_{\alpha}\ ,   
\end{align}
where $q^{\prime}= u,d,s,c$.
{}Finally, the electroweak penguin operators arise from the  
two Feynman diagrams represented 
in Fig.~\ref{figO2} (left) and Fig.~\ref{figO2} (right) where
$Z,\gamma$ is exchanged from a quark line and 
from the $W$ line, respectively. They have the
following expressions:  
\begin{align}\label{eq32}
O_{7}& =\frac{3}{2}\bar{q} \gamma_{\mu}(1-\gamma{_5})b \sum_{q'}e_{q^{\prime}}
\bar{q}^{\prime} \gamma^{\mu}(1+\gamma{_5})q^{\prime}\ , \nonumber \\ 
O_{8}& =\frac{3}{2}\bar{q}_{\alpha} \gamma_{\mu}(1-\gamma{_5})b_{\beta} 
\sum_{q'}e_{q^{\prime}}\bar{q}^{\prime}_{\beta}\gamma^{\mu}(1+\gamma{_5})
q^{\prime}_{\alpha}\ , \nonumber \\
O_{9}& =\frac{3}{2}\bar{q} \gamma_{\mu}(1-\gamma{_5})b \sum_{q'}e_{q^{\prime}}
\bar{q}^{\prime} \gamma^{\mu}(1-\gamma{_5})q^{\prime}\ , \nonumber \\
 O_{10}& =\frac{3}{2}\bar{q}_{\alpha} \gamma_{\mu}(1-\gamma{_5})b_{\beta} 
\sum_{q'}e_{q^{\prime}}\bar{q}^{\prime}_{\beta}\gamma^{\mu}(1-\gamma{_5})q^{\prime}_{\alpha}\ ,
\end{align}
where  $e_{q^{\prime}}$ denotes the  electric  charge of $q^{\prime}$. 
%
%
\subsubsection{Wilson coefficients}\label{sec4.1.2}
%
As we  mentioned in the preceding  subsection,  the Wilson 
coefficients~\cite{ref9}, $C_{i}(\mu)$,  represent the 
physical contributions from scales higher than $\mu$ (the OPE describes
physics for scales lower than $\mu$).
Since  QCD has the property of  asymptotic freedom, they can 
be calculated in perturbation theory. The Wilson coefficients include
the contributions of all heavy particles,
such as the top quark, the $W$ bosons, and the charged Higgs boson. Usually,
the scale $\mu$ is chosen to be of  
 ${\cal O}(m_{b})$ for $B$ decays. The Wilson coefficients have been  
calculated to next-to-leading order (NLO).
The evolution of $C(\mu)$ (the matrix that includes $C_{i}(\mu)$) is given by,
\begin{equation}\label{eq33}
C(\mu)= U(\mu,M_{W})C(M_{W})\ , 
\end{equation}
where $U(\mu,M_{W})$  is the QCD evolution matrix:
\begin{equation}\label{eq34}
U(\mu,M_{W})= \biggl[ 1+ \frac{\alpha_{s}(\mu)}{4 \pi}J \biggr] U^{0}(\mu,M_{W}
) \biggl[1- \frac{\alpha_{s}(M_{W})}{4 \pi}J \biggr] \ ,
\end{equation}
with $J$ the matrix  summarizing the next-to-leading order corrections and
$U^{0}(\mu,M_{W})$ the evolution matrix in the leading-logarithm  approximation.
Since the strong interaction is independent of
quark flavor, the $C(\mu)$ are the same for all $B$ decays. At the scale 
$\mu=m_{b}=5$ GeV,  $C(\mu)$ take   the  values summarized in Table~\ref{tab1}.
To be consistent, the matrix elements of the operators, $O_{i}$, should also
be renormalized to the one-loop 
order. This results in the effective Wilson coefficients, 
$C_{i}^{\prime}$, which satisfy the constraint,
\begin{eqnarray}\label{eq35}
C_{i}(m_{b})\langle O_{i}(m_{b})\rangle=C_{i}^{\prime}{\langle O_{i}\rangle}^{tree}\ , 
\end{eqnarray}
here ${\langle O_{i}\rangle}^{tree}$ are the matrix elements 
at the tree level. These matrix elements 
will be evaluated 
within the factorization approach. From Eq.~(\ref{eq35}), the relations between
$C_{i}^{\prime}$ and $C_{i}$ are~\cite{ref10,ref11}:
\begin{align}\label{eq36}
C_{1}^{\prime}& =C_{1}\ ,\; \nonumber &
C_{2}^{\prime}& =C_{2}\ , \nonumber \\
C_{3}^{\prime}& =C_{3}-P_{s}/3\ ,\; \nonumber &
C_{4}^{\prime}& =C_{4}+P_{s}\ , \nonumber \\
C_{5}^{\prime}& =C_{5}-P_{s}/3\ ,\; \nonumber &
C_{6}^{\prime}& =C_{6}+P_{s}\ , \nonumber \\
C_{7}^{\prime}& =C_{7}+P_{e}\ ,\; \nonumber &
C_{8}^{\prime}& =C_{8}\ , \nonumber \\
C_{9}^{\prime}& =C_{9}+P_{e}\ ,\;  & 
C_{10}^{\prime}& =C_{10}\ ,
\end{align}
where,
\begin{align}\label{eq37}
P_{s} & =(\alpha_{s}/8\pi)C_{2}(10/9+G(m_{c},\mu,q^{2}))\ , \nonumber \\
P_{e} & =(\alpha_{em}/9\pi)(3C_{1}+C_{2})(10/9+G(m_{c},\mu,q^{2}))\ , 
\end{align}
and
\begin{eqnarray}\label{eq38}
 G(m_{c},\mu,q^{2})=4\int_{0}^{1}dx \ x(x-1){\rm ln} \frac{m_{c}^{2}-x(1-x)q^{2}}{\mu^{2}}\ .
\end{eqnarray}
Here $q^{2}$ is the typical momentum transfer of the gluon or 
photon in the penguin diagrams and 
$G(m_{c},\mu,q^{2})$ 
has the following explicit expression~\cite{ref12}, 
\begin{align}\label{eq39}
&\mathscr{R}\!e\;  G  = \frac{2}{3} \left({\rm ln} \frac{m_{c}^{2}}{\mu^{2}}-
  \frac{5}{3}-4 \frac{m_{c}^{2}}{q^{2}}+
\left(1+2\frac{m_{c}^{2}}{q^{2}}\right)\sqrt{1-4\frac{m_{c}^{2}}{q^{2}}}{\rm ln} \frac{1+\sqrt{1-4
\frac{m_{c}^{2}}{q^{2}}}}{1-\sqrt{1-4\frac{m_{c}^{2}}{q^{2}}}}\right), \nonumber   \\
&\mathscr{I}\!m \; G  = -\frac{2}{3}\left(1+2\frac{m_{c}^{2}}{q^{2}}\right)\sqrt{1-4\frac{m_{c}^{2}}{q^{2}}}\ . 
\end{align}
Based on simple arguments at the quark level, 
the value of $q^{2}$ is chosen in the range
$0.3 < q^{2}/m_{b}^{2} < 0.5$~\cite{Gardner:1997yx,ref3}. 
{}From Eqs.~(\ref{eq36}-\ref{eq39}) 
we can obtain numerical values for  
 $C_{i}^{\prime}$. These values are listed in  Table~\ref{tab2}, where we have 
taken $\alpha_{s}(m_{Z})=0.112, \;   
\alpha_{em}(m_{b})=1/132.2,\;  m_{b}=5$ GeV,
and $ \; m_{c}=1.35$ GeV.
%
\subsubsection{Effective  Hamiltonian}\label{sec4.1.3}
%
In any phenomenological treatment of the weak decays of hadrons, 
the starting point is the weak effective
Hamiltonian at low energy~\cite{ref13}. It is
obtained by integrating out the heavy fields (e.g. the top quark, $W$ and $Z$
bosons) from the standard model
Lagrangian. It can be written as:
%
\begin{equation}\label{eq40}
{\cal H}_{eff}=\frac {G_{F}}{\sqrt 2} \sum_{i} V_{CKM} C_{i}(\mu)O_i(\mu)\ ,
\end{equation}
where $G_{F}$ is the Fermi constant, $V_{CKM}$ is the 
CKM matrix element (see Section~\ref{sec4.3}),
$C_{i}(\mu)$ are
the Wilson coefficients (see Section~\ref{sec4.1.2}),
$O_i(\mu)$ are the operators from the operator product expansion
(see Section~\ref{sec4.1.1}), and $\mu$ represents the renormalization scale.
We emphasize that the amplitude corresponding
to the effective Hamiltonian for a given decay 
is independent of the scale $\mu$. 
In the present case, since we analyze direct  $CP$  violation in 
$B$ decays, we take into account both tree and penguin diagrams.
{}For the penguin diagrams, we include all operators $O_{3}$ to $O_{10}$. 
Therefore, the effective Hamiltonian used will be,
%
\begin{equation}\label{eq41}
{\cal H}_{eff}^{\bigtriangleup B=1}=\frac {G_{F}}{\sqrt 2} \biggl[ V_{ub}V_{us}^{\ast}(C_{1}O_{1}^{s} + 
C_{2}O_{2}^{s})- V_{tb}V_{ts}^{\ast} \sum_{i=3}^{10} C_{i}O_{i} \biggr] + H.c.\ ,
\end{equation}
%
and consequently, the decay amplitude can be expressed as  follows, 
\begin{multline}\label{eq42}
A(B \rightarrow V_1 V_2) =
\frac {G_{F}}{\sqrt 2} \biggl[  V_{ub}V_{us}^{\ast}\bigl( C_{1}\langle V_1 V_2 | O_{1}^{s}| B \rangle + 
C_{2}\langle V_1 V_2 |O_{2}^{s}| B \rangle \bigr) - \\
 V_{tb}V_{ts}^{\ast} \sum_{i=3}^{10} C_{i}\langle V_1 V_2 |O_{i}| B \rangle \biggr]+ H.c.\ ,
\end{multline}
where $\langle V_1 V_2 |O_{i}| B \rangle$ are the hadronic matrix elements.  
They  describe the transition
between the initial
state and the final state for scales lower than $\mu$ and include, 
up to now, the main uncertainties
in the calculation since they involve non-perturbative effects.
%
\subsection{New expression of helicity amplitudes $\boldsymbol{h_{ij}}$ according to Wilson Coefficients}\label{sec4.2}
%
\subsubsection{General helicity amplitude}\label{sec4.2.1}
%
The weak hadronic matrix element is expressed as the  
sum of three helicity matrix elements, each of which 
takes the form 
$ H_{\lambda}\bigl(B \rightarrow {{\rho}^0}(\omega) V_2 \bigr) 
= \langle V_1 V_2 | {H_w}^{eff} | B \rangle $,  
and is defined by gathering all the Wilson 
coefficients of both the tree and penguin operators. 
Linear combinations of those coefficients arise, such as 
$c_{t_i}^{V_i}$ (tree diagram contribution) 
and $ c_{p_i}^{V_i}$ (penguin diagram 
contribution). Then, in the case of $B \rightarrow {{\rho}^0}(\omega) V_2 $,
($V_1= \rho^0$ or $\omega$), the 
helicity amplitude $H_ {\lambda}\bigl(B \rightarrow {{\rho}^0}(\omega) V_2
\bigr)$
has the general following expression:
\begin{multline}\label{eq43}
H_{\lambda}\bigl(B \rightarrow {{\rho}^0}(\omega) V_2 \bigr)
=\Big(V_{ub}V_{us}^{*}c_{t_{1}}^{\rho}-
V_{tb}V_{ts}^{*}c_{p_{1}}^{\rho}\Big)\bigg\lbrace 
\beta_{1}^{\rho}\varepsilon_{\alpha \beta \gamma
  \delta}\epsilon_{V_2}^{*\alpha}(\lambda)\epsilon_{\rho}^{* \beta}
(\lambda)P_{B}^{\gamma}P_{V_2}^{\delta} \\
+i\Big(\beta_{2}^{\rho}\epsilon_{V_2}^{*}(\lambda)\epsilon_{\rho}^{*}(\lambda)
- \beta_{3}^{\rho}(\epsilon_{V_2}^{*}(\lambda)\sdot
P_{B})
(\epsilon_{\rho}^{*}(\lambda) \sdot P_{B})\Big)\bigg\rbrace 
 + \Big(V_{ub}V_{us}^{*}c_{t_{2}}^{\rho}-V_{tb}V_{ts}^{*}c_{p_{2}}^{\rho}\Big) \\ 
\bigg\lbrace \beta_{4}^{\rho}\varepsilon_{\alpha \beta \gamma \delta}
\epsilon_{\rho}^{* \alpha}(\lambda)\epsilon_{V_2}^{* \beta}
(\lambda)P_{B}^{\gamma}P_{\rho}^{\delta}
+i\Big(\beta_{5}^{\rho}\epsilon_{\rho}^{*}(\lambda)\epsilon_{V_2}^{*}(\lambda)
- \beta_{6}^{\rho}(\epsilon_{\rho}^{*}
(\lambda) \sdot P_{B})(\epsilon_{V_2}^{*}(\lambda) \sdot
P_{B})\Big)\bigg\rbrace \\
+ \frac{\tilde{\Pi}_{\rho
  \omega}}{{(s_{\rho}-m_{\omega}^{2})+im_{\omega}\Gamma_{\omega}}} 
\Biggl[ \Big(V_{ub}V_{us}^{*}c_{t_{1}}^{\omega}-
V_{tb}V_{ts}^{*}c_{p_{1}}^{\omega}\Big)\bigg\lbrace 
\beta_{1}^{\omega}\varepsilon_{\alpha \beta \gamma
  \delta}\epsilon_{V_2}^{*\alpha}(\lambda)\epsilon_{\omega}^{* \beta}
(\lambda)P_{B}^{\gamma}P_{V_2}^{\delta} \\
+i\Big(\beta_{2}^{\omega}\epsilon_{V_2}^{*}(\lambda)\epsilon_{\omega}^{*}(\lambda)
- \beta_{3}^{\omega}(\epsilon_{V_2}^{*}(\lambda)\sdot
P_{B})
(\epsilon_{\omega}^{*}(\lambda) \sdot P_{B})\Big)\bigg\rbrace 
 + \Big(V_{ub}V_{us}^{*}c_{t_{2}}^{\omega}-V_{tb}V_{ts}^{*}c_{p_{2}}^{\omega}\Big) \\ 
\bigg\lbrace \beta_{4}^{\omega}\varepsilon_{\alpha \beta \gamma \delta}
\epsilon_{\omega}^{* \alpha}(\lambda)\epsilon_{V_2}^{* \beta}
(\lambda)P_{B}^{\gamma}P_{\omega}^{\delta}
+i\Big(\beta_{5}^{\omega}\epsilon_{\omega}^{*}(\lambda)\epsilon_{V_2}^{*}(\lambda)
- \beta_{6}^{\omega}(\epsilon_{\omega}^{*}
(\lambda) \sdot P_{B})(\epsilon_{V_2}^{*}(\lambda) \sdot
P_{B})\Big)\bigg\rbrace \Biggr] \ ,
\end{multline}
with $\epsilon_{V_2, \rho, \omega}(\lambda)$ being the $K^*$,  $\rho^{0}$ and
$\omega$ polarization vectors expressed 
in the $B$ rest frame. Finally $\varepsilon_{\alpha \beta \gamma \delta}$ 
is the  antisymmetric tensor in 
Minkowski space.

In Eq.~(\ref{eq43}) the parameters $\beta_{i}$ are mainly 
the form factors describing
transitions between vector mesons. They take the form:
\begin{align}\label{eq44} 
\beta_{1,4}^{V_1} & =\frac{G_{F}}{2} f_{V_1,V_2}m_{V_1,V_2}\frac{2}{m_{B}+m_{V_2,V_1}}V^{B\to V_2,
V_1}(m^{2}_{V_1,V_2})\ ,  \\  
\beta_{2,5}^{V_1} & =\frac{G_{F}}{2} f_{V_1,V_2}m_{V_1,V_2}(m_{B}+m_{V_2,V_1})A_{1}^{B\to V_2,V_1}
(m^{2}_{V_1,V_2})\ ,    \\
\beta_{3,6}^{V_1} & =\frac{G_{F}}{2} f_{V_1,V_2}m_{V_1,V_2}\frac{2}{m_{B}+m_{V_2,V_1}}A_{2}^{B\to V_2,
V_1}(m^{2}_{V_1,V_2})\ ,
\end{align}
here $f_{V_1,V_2}$ is either the $\rho^{0}, \omega$ 
or $K^{*}$ decay constant.
$V^{B\to V_2,V_1}$ and $A_{i}^{B \to V_2,V_1}$ 
are respectively the vector and axial form
factors defined in Eqs.~(\ref{eq14}-\ref{eq16}).
It is worth noticing that the tensorial terms which enter 
$H_{\lambda}\bigl(B \rightarrow \rho^{0}(\omega) V_2 \bigr)$ 
become simplified in the $B$
rest frame because the four-momentum of the $B$ 
is given by $P_B = {(m_B, {\vec 0})}$. Then, using the
orthogonality properties of ${\epsilon}_{V_i}{(\lambda)}$, 
the helicity amplitude 
$H_{\lambda}\bigl(B \rightarrow \rho^{0}(\omega) V_2 \bigr)$ acquires a 
much simpler expression than above:
\begin{multline}\label{eq45} 
H_{\lambda}\bigl(B \rightarrow \rho^{0}(\omega) V_2 \bigr) = 
iB^{\rho}_\lambda(V_{ub}V_{us}^{*}c_{t_1}^{\rho}-V_{tb}V_{ts}^{*}c_{p_1}^{\rho})+
iC^{\rho}_\lambda(V_{ub}V_{us}^{*}c_{t_2}^{\rho}-V_{tb}V_{ts}^{*}c_{p_2}^{\rho}) +\\ 
\frac{\tilde{\Pi}_{\rho
  \omega}}{(s_{\rho}-m_{\omega}^{2})+im_{\omega}\Gamma_{\omega}}
\Bigl[ iB^{\omega}_\lambda(V_{ub}V_{us}^{*}c_{t_1}^{\omega}-V_{tb}V_{ts}^{*}c_{p_1}^{\omega})+
iC^{\omega}_\lambda(V_{ub}V_{us}^{*}c_{t_2}^{\omega}-V_{tb}V_{ts}^{*}c_{p_2}^{\omega})\Bigr]
\ ,
\end{multline}
where the terms $B_\lambda^{V_1}$ and $C_\lambda^{V_1}$ 
take the following forms for the helicity
($\lambda$) values, $-1,0,+1$:
\begin{gather}\label{eq46}  
B^{V_1}_{\lambda=0}=\beta_{2}^{V_1}\frac{m_{B}^2 -(m_{V_2}^2 + m_{V_1}^2 )}{2m_{V_2}m_{V_1}} - 
\beta_{3}^{V_1}\frac{|\vec{p}|^{2}m_{B}^2 }
{m_{V_2}m_{V_1}}\ , \\
C^{V_1}_{\lambda=0}=\beta_{5}^{V_1}\frac{m_{B}^2 -(m_{V_2}^2 + m_{V_1}^2 )}{2m_{V_2}m_{V_1}} - 
\beta_{6}^{V_1}\frac{|\vec{p}|^{2}m_{B}^2}
{m_{V_2}m_{V_1}}\ , \\
B^{V_1}_{\lambda=\pm{1}} = \mp \beta_{1}^{V_1}m_{B}|\vec{p}| - \beta_{2}^{V_1}\ , \\
C^{V_1}_{\lambda=\pm{1}} = \mp \beta_{4}^{V_1}m_{B}|\vec{p}| - \beta_{5}^{V_1}\ . 
\end{gather}
In the above equations, $|\vec{p}|$ is the momentum common to the $V_1$ and
$V_2$ particles in the $B$ rest frame. It takes the form:
\begin{equation}\label{eq47}  
|\vec{p}|=\frac{ \sqrt{ [m_{B}^{2}-(m_{V_1}+m_{V_2})^{2}][m_{B}^{2}-(m_{V_1}-m_{V_2})^{2}]}}{2m_{B}}\ ,
\end{equation}
where $m_{1}$ and $m_{2}$ are the vector masses.
Taking into account the previous relations, we arrive at the final form for
the amplitudes  $H_{\lambda}\Bigl(B \rightarrow \rho^{0}(\omega) V_2 \Bigr)$:
\begin{multline}\label{eq48}
H_{\lambda=^0_{\pm{1}}} \Bigl(B \rightarrow \rho^{0}(\omega) V_2 \Bigr)= \\
 A\lambda^{2} \Bigg \lbrace \bigg \lbrack 
R_1^{\rho} B^{\rho}_{\lambda=^0_{\pm{1}}}+ R_2^{\rho}C^{\rho}_{\lambda=^0_{\pm{1}}}\bigg \rbrack 
+ \ i\ \bigg \lbrack I_1^{\rho} B^{\rho}_{\lambda=^0_{\pm{1}}} + I_2^{\rho}
C^{\rho}_{\lambda=^0_{\pm{1}}} \bigg \rbrack \Bigg \rbrace \hspace{2.7cm} \\
+ \frac{
\tilde{\Pi}_{\rho
 \omega}}{(s_{\rho}-m_{\omega}^{2})+im_{\omega}\Gamma_{\omega}}
\Biggl[
A\lambda^{2} \Bigg \lbrace \bigg \lbrack R_1^{\omega}
B^{\omega}_{\lambda=^0_{\pm{1}}} + R_2^{\omega} C^{\omega}_{\lambda=^0_{\pm{1}}}\bigg \rbrack \\
+ \ i\ \bigg \lbrack I_1^{\omega}  B^{\omega}_{\lambda=^0_{\pm{1}}} + I_2^{\omega}
C^{\omega}_{\lambda=^0_{\pm{1}}} \bigg \rbrack \Bigg \rbrace
\Biggr]\ ,
\end{multline}
where one defines,
\begin{align}\label{eq48a}
R_i^{V_1} & = \eta \lambda^{2} c_{t_{i}}^{V_1} - \mathscr{I}\!m{(c_{p_{i}}^{V_1})}\ , \\
I_i^{V_1} & = \rho \lambda^{2} c_{t_{i}}^{V_1} + \mathscr{R}\!e{(c_{p_{i}}^{V_1})}\ ,
\end{align}
with $V_1$ being either $\rho^0$ or $\omega$.
From Eq.~(\ref{eq48}), the density-matrix elements $h_{{\lambda},
  {\lambda}'}$ can be derived automatically and on has:
\begin{equation}\label{eq49}
h_{{\lambda}, {\lambda}'} = H_{\lambda}\bigl(B \rightarrow \rho^{0}(\omega) V_2 \bigr) H^{*}_{\lambda'}\bigl(B \rightarrow \rho^{0}(\omega) V_2 \bigr)\ . 
\end{equation}
Because of the  hermiticity of the matrix ($h_{{\lambda}, {\lambda}'}$), 
only six elements must be calculated.
%
\subsubsection{Explicit amplitudes for the $\boldsymbol{B}$ decays investigated}\label{sec4.2.2}
%
By applying the formalism described in Section III, one gets in the case of
the $\rho^{0}$ production, the following linear 
combinations of the effective Wilson coefficients:
\newline
\noindent for the decay ${\bar B}^{0} \rightarrow {\bar K}^{* 0} \rho^{0}$: 
\begin{align}\label{eq50a}
c_{t_{1}}^{\rho} & = C^{\prime}_{1} + \frac{C^{\prime}_{2}}{N_{c}}\ , 
& c_{p_{1}}^{\rho}  =\frac{3}{2} \Bigl(C^{\prime}_{9} + \frac{C^{\prime}_{10}}{N_{c}}+
C^{\prime}_{7}+\frac{C^{\prime}_8}{N_{c}}\Bigr)\ , \hspace{2.em} \nonumber \\
c_{t_{2}}^{\rho} & = 0\ , 
& c_{p_{2}}^{\rho}  =-\Bigl(C^{\prime}_4 +
\frac{C^{\prime}_3}{N_{c}}\Bigr)+\frac{1}{2}\Bigl(C^{\prime}_{10}+\frac{C^{\prime}_{9}}{N_{c}}\Bigr)\ ,
\end{align}
where $C^{\prime}_{i}$ are listed in Table 2.
The coefficients, $c_{t_{i}}^{\rho}$, relate to the tree diagrams
and $c_{p_{i}}^{\rho}$  
to the penguin diagrams. 
To simplify the formulas we used $N_c$ for $N_{c}^{eff}$ in
the expressions (Eqs.~(\ref{eq50a})-(\ref{eq50d})). 
\newline
\noindent for the decay $B^{-} \rightarrow K^{* -} \rho^{0}$:
\begin{align}\label{eq50b}
c_{t_{1}}^{\rho} & = C^{\prime}_{1} + \frac{C^{\prime}_{2}}{N_{c}}\ , 
& c_{p_{1}}^{\rho}  =\frac{3}{2} \Bigl(C^{\prime}_{9} + \frac{C^{\prime}_{10}}{N_{c}}+
C^{\prime}_{7}+\frac{C^{\prime}_8}{N_{c}}\Bigr)\ , \nonumber \\
c_{t_{2}}^{\rho} & = C^{\prime}_{2} + \frac{C^{\prime}_{1}}{N_{c}}\ , 
& c_{p_{2}}^{\rho}  =C^{\prime}_4 +
\frac{C^{\prime}_3}{N_{c}}+C^{\prime}_{10}+\frac{C^{\prime}_{9}}{N_{c}}\
. \hspace{0.8cm}
\end{align}

In the case of $\omega$
production one obtains the following linear combinations of 
effective Wilson coefficients:
\newline
\noindent for the decay ${\bar B}^{0} \rightarrow {\bar K}^{* 0} \omega$:
\begin{align}\label{eq50c}
c_{t_{1}}^{\omega} & = 0\ , 
& c_{p_{1}}^{\omega}  = -C^{\prime}_{4}-\frac{C^{\prime}_3}{N_{c}}+\frac{1}{2}
\Bigl(C^{\prime}_{10} + \frac{C^{\prime}_{9}}{N_{c}}\Bigr)\ , \hspace{4.4cm} \nonumber \\
c_{t_{2}}^{\omega} & = C^{\prime}_{1} + \frac{C^{\prime}_{2}}{N_{c}}\ , 
& c_{p_{2}}^{\omega}  =2\Bigl(C^{\prime}_3 +
\frac{C^{\prime}_4}{N_{c}}+ C^{\prime}_5 + \frac{C^{\prime}_6}{N_{c}}\Bigr) +
\frac{1}{2} \Bigl(C^{\prime}_{9}+\frac{C^{\prime}_{10}}{N_{c}} 
+ C^{\prime}_{7}+\frac{C^{\prime}_{8}}{N_{c}} \Bigr)\ .
\end{align}
\newline
\noindent for the decay $B^{-} \rightarrow K^{* -} \omega$:
\begin{align}\label{eq50d}
c_{t_{1}}^{\omega} & = C^{\prime}_{2} + \frac{C^{\prime}_{1}}{N_{c}}\ , 
& c_{p_{1}}^{\omega}  = C^{\prime}_{4}+\frac{C^{\prime}_3}{N_{c}}+
\Bigl(C^{\prime}_{10} + \frac{C^{\prime}_{9}}{N_{c}}\Bigr)\ , \hspace{5.1cm} \nonumber \\
c_{t_{2}}^{\omega} & = C^{\prime}_{1} + \frac{C^{\prime}_{2}}{N_{c}}\ , 
& c_{p_{2}}^{\omega}  =2\Bigl(C^{\prime}_3 +
\frac{C^{\prime}_4}{N_{c}}+ C^{\prime}_5 + \frac{C^{\prime}_6}{N_{c}}\Bigr) +
\frac{1}{2} \Bigl(C^{\prime}_{9}+\frac{C^{\prime}_{10}}{N_{c}} 
+ C^{\prime}_{7}+\frac{C^{\prime}_{8}}{N_{c}} \Bigr)\ .
\end{align}
We refer to Appendix A for details of the helicity amplitudes, 
while for  
the channel $B^{\pm} \rightarrow \rho^{0}(\omega) \rho^{\pm}$ 
we refer to Appendix B.
%
\subsection{CKM matrix and form factors}\label{sec4.3}
In phenomenological applications, the widely used representation  
of the CKM matrix is
the {\it Wolfenstein parametrization}~\cite{ref15}.
In this approach, the four independent parameters are $\lambda_c, A, \rho$ and
 $\eta$. Then, by expanding each element of the
matrix as a power series in the parameter 
$\lambda_c = \sin \theta_{c} = 0.2209$ ($\theta_{c}$ is
the Gell-Mann-Levy-Cabibbo angle), one finds ($O(\lambda_c^4)$ is neglected)
\begin{equation}\label{eq51}
{\hat V}_{CKM}= \left( \begin{array}{ccc}
1-\frac{1}{2} \lambda_c^{2} &  \lambda_c                    & A\lambda_c^{3}(\rho-i\eta) \\
-\lambda_c                  & 1-\frac{1}{2}\lambda_c^{2}    & A\lambda_c^{2}             \\
A\lambda_c^{3}(1-\rho-i\eta)& -A\lambda_c^{2}               &      1                   \\
\end{array}  \right)\ ,
\end{equation}
where $\eta$  plays the role of the  $CP$-violating phase. In this
parametrization, even though it is an approximation
in $\lambda_c$, the CKM matrix satisfies unitarity exactly, which means,
\begin{equation}\label{eq52}
{\hat V}_{CKM}^{\dagger} \cdot {\hat V}_{CKM} = {\hat I} = {\hat V}_{CKM}
\cdot  {\hat V}_{CKM}^{\dagger}\ .
\end{equation}
The form factors, $V(k^{2})$ and $A_{j}(k^{2})$,   
depend on the inner structure of the 
hadrons. Here we will adopt two different theoretical approaches. The first
was proposed by Bauer, Stech,
and Wirbel~\cite{ref20} (BSW), who used the overlap 
integrals of wave functions in order to evaluate 
the meson-meson matrix elements of 
the corresponding current. In that case the momentum  
dependence of the form factors is based on a single-pole ansatz.
The second approach was developed by Guo and Huang (GH)~\cite{ref22},
who modified the BSW model by using some wave functions 
described in the light-cone framework. 
Nevertheless, both of these models use phenomenological form factors 
which are parametrized by making the assumption of 
nearest pole dominance. 
The explicit $k^{2}$ dependence of the form factor 
is~\cite{ref20,ref25}:
\begin{gather}\label{eq53}
V(k^{2})=\frac{h_{V}}{\left( 1-\frac{k^{2}}{m_{V}^{2}} \right)}\ , \;\;\;  \;\;\; 
A_{j}(k^{2})=\frac{h_{A_{j}}}{\left( 1-\frac{k^{2}}{m_{A_{j}}^{2}}\right)}\ , 
\end{gather} 
where $m_{A_{j}}$ and $m_{V}$ are the pole masses 
associated with the transition current  and  
$h_{V}$ and $h_{A_{j}}$
are the values of the form factors at $q^{2}=0$.
%
\section{Monte-Carlo simulations: computation of $\boldsymbol{h_{ij}}$ and general results}\label{sec5}
%
\subsection{Numerical inputs}\label{sec5.1}
%
%
\subsubsection{CKM values}\label{sec5.1.1}
%
In our numerical calculations  we have several parameters: $q^{2},
N_{c}^{eff}$ and the CKM matrix elements 
in the Wolfenstein parametrization. 
As mentioned in Section IV, the value of $q^{2}$ is conventionally
chosen to be in the range $0.3<q^{2}/{m_{b}}^{2}<0.5$. 
The CKM matrix, which should be determined from
experimental data,  is expressed  in terms  of the 
Wolfenstein parameters, $ A,\; \lambda_c,\; 
\rho$, and $\eta $~\cite{ref15}.
Here, we shall use the latest values~\cite{ref26}, which were extracted from
charmless semileptonic $B$ decays, ($|V_{ub}|$),  
charmed semileptonic $B$ decays,  ($|V_{cb}|$),
$s$ and $d$ mass oscillations, $\Delta m_{s}, \Delta m_{d}$,
and $CP$ violation in the kaon system ($\epsilon_{K}$), ($\rho, \eta$). 
Hence, one has,
\begin{equation}\label{eq54}
\lambda_c=0.2237\ , \;\; A=0.8113\ , \;\;  0.190 < \rho < 0.268\ , \;\; 0.284< \eta <0.366\ .
\end{equation}
These values  respect the unitarity triangle as well (see also
Table~\ref{tab3}). In our numerical simulations, we will use the average
values of $\rho$ and $\eta$.
%
\subsubsection{Quark masses}\label{sec5.1.2}
%
The running quark masses are used in order to calculate the 
matrix elements of penguin operators. 
The quark mass is evaluated at the scale $\mu \simeq m_{b}$ in 
$B$ decays. Therefore one has~\cite{ref27},
\begin{align}\label{eq55}
m_{u}(\mu=m_{b})& = 2.3 \;{\rm MeV}\ , & m_{d}(\mu=m_{b})& = 4.6 \;{\rm MeV}\ , \nonumber\\
m_{s}(\mu=m_{b})& = 90 \;{\rm MeV}\ , &  m_{b}(\mu=m_{b})& = 4.9 \;{\rm GeV}\ ,
\end{align}
which corresponds to $m_{s}(\mu= 1\;{\rm GeV}) 
= 140 \;{\rm MeV}$. For meson  masses,
we shall use the following values~\cite{ref16}:
\begin{align}\label{eq56}
 m_{B^{\pm}}& = 5.279 \; {\rm GeV}\ , & m_{K^{*}0}& = 0.896 \; {\rm GeV}\ ,
&  m_{\omega}& = 0.782 \; {\rm GeV}\ , \nonumber\\
m_{B^{0}}&   = 5.279 \; {\rm GeV}\ , & m_{\rho^{\pm}}& = 0.770 \; {\rm GeV}\ , 
& m_{\pi^{\pm}}& = 0.139 \; {\rm GeV}\ ,  \nonumber\\
m_{K^{* \pm}}& = 0.892 \; {\rm GeV}\ , & m_{\rho^{0}}& = 0.770 \; {\rm GeV}\ , 
& m_{\pi^{0}}&  = 0.135 \; {\rm GeV}\ .
\end{align}
%
%
\subsubsection{Form factors and decay constants}\label{sec5.1.3}
%
In Table~\ref{tab4} we list the relevant form factor values at zero 
momentum transfer~\cite{ref20,ref22,ref28}  for the
$B \rightarrow K^{*}$,  $B \rightarrow \rho$ and $B \rightarrow \omega$ transitions.
The different models are defined as follows~: model (1) is  the BSW model
where the $q^{2}$ dependence of the 
{}form factors is described  by  a single-ansatz. Model (2)
is the GH model with the 
same momentum dependence as model (1). Finally, 
we define the decay constant for vector ($f_{V}$) meson as usual by,
\begin{align}\label{eq57}
\sqrt{2} \langle \rho(q) | \bar{q}_{1} \gamma_{\mu} q_{2} | 0 \rangle & =
f_{\rho} m_{\rho} \epsilon_{\rho}
 \;\; {\rm for}\;  \rho \; {\rm \; and \; otherwise}\ , \nonumber \\
\langle V(q) | \bar{q}_{1} \gamma_{\mu} q_{2} | 0 \rangle & = f_{V} m_{V} \epsilon_{V}\ ,
\end{align}
with $q$ being the momentum of the vector meson and  
$m_{V}$ and $\epsilon_{V}$ being the mass
and polarization vector of the vector meson, respectively.
Numerically, in our calculations, we  take~\cite{ref16},
\begin{gather}\label{eq58}
   f_{K^{*}} = 214 \; {\rm MeV}\ ,  \; 
f_{\rho} = 221 \; {\rm MeV}\ ,  \;  f_{\omega}  = 195 \; {\rm MeV}\ . 
\end{gather}
{}Finally, the free parameter 
(effective number of color, $N_c^{eff}$) is taken to lie between the
lower(upper) limits 0.66(2.84) for $b \rightarrow s$ transition. Nevertheless,
we focus our analysis on values of $N_c^{eff}$ bigger than 1, as  
suggested in~\cite{refolive}. Regarding the $b \rightarrow d$ transition, the
lower(upper) limits for $N_c^{eff}$ are 0.98(2.01)~\cite{refolive}.
%
\subsection{Simulation of the $\boldsymbol{{\rho}^0 - \omega}$ mixing}\label{sec5.2}
\begin{figure}
\centering\includegraphics[height=14.cm, width = 12.cm,clip=true]{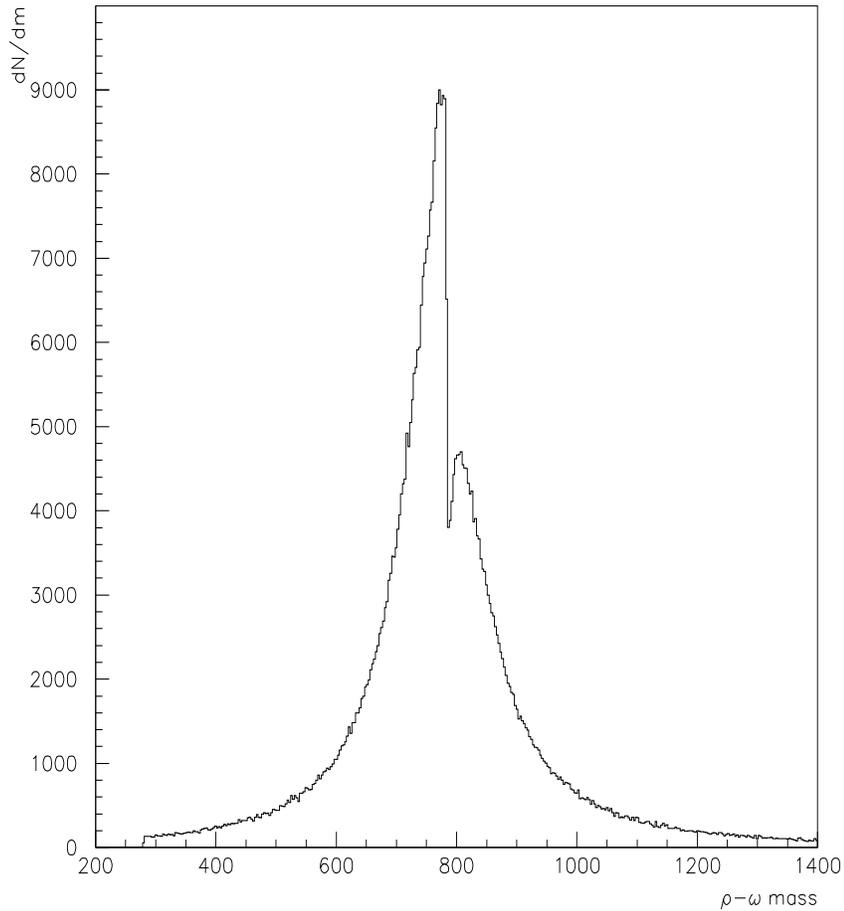}
\caption{Spectrum of ${\rho^{0}}- {\omega}$ mixing (in MeV/$c^2$), simulated 
by the interference of two Breit-Wigner curves.}
\label{fig1}
\end{figure}

%
All the channels studied here include at least one ${\rho}^0$ meson 
which mixes with the $\omega$ meson.
The other vector mesons are either a $K^{* 0,\pm}$ or a ${\rho}^{\pm}$. 
Thus, the mass of each resonance
is generated according to a relativistic Breit-Wigner:
\begin{equation}\label{eq59}
\frac{d\sigma}{dM^2}  =   {C_N}   \frac{\Gamma_R M_R}{{(M^2-{M^2_R})}^2 +
  {(\Gamma_R M_R)}^2}\ , 
\end{equation}
where $C_N$ is a normalization constant. 
In Eq.~(\ref{eq59}), $M_R$ and $\Gamma_R$ are
respectively the mass and the width of the vector meson which have been
determined experimentally. $M$ is the mass of the generated resonance.
A simple and phenomenological relation describing the amplitude for   
${\rho}^0 - \omega$ mixing is used for the Monte-Carlo 
simulations~\cite{LANGACKER}. In the expression for  
the Breit-Wigner, the ${\rho}^0$-propagator is replaced by the following one:
 \begin{eqnarray}\label{eq60}
 \frac{1}{s_{\rho \omega}}= {\frac{1}{s_{\rho}}}+{\frac{T_{\omega}}{T_ {\rho}}}
 {\frac{{\Pi}_{\rho \omega}}{{s_{\rho}}{s_{\omega}}}}\ ,
\end{eqnarray}
where $T_{\omega}$ and $T_{\rho}$ are respectively the $\omega$ and $\rho$
production amplitudes. In addition, ${\tilde{\Pi}}_{\rho \omega}$ 
is the mixing parameter for which
recent values come from $e^+ e^- \rightarrow \pi^+ \pi^-$ annihilations. 
Explanations have been
already been given in Section III. Finally, $1/s_V$ 
has the same definition as in Eq.~(\ref{eq22}).
Because the same physical processes enter the production of both the
${\rho}^0$ and $\omega$ 
resonances (they are both made out from $u {\bar u}$ and $d {\bar d}$ quark
pairs with the same weight $1/2$), it seems
natural to choose ${T_{\omega}}/{T_ {\rho}} = 1 $. 
So, the invariant mass distribution of the $\pi^+ \pi^-$ system 
becomes simplified, being given by,
\begin{equation}\label{eq61}
{d\sigma}/{dm^2}  \propto   {\bigl|{\cal A}({{\rho^{0}} ({\omega})})\bigr|}^2\ , 
\end{equation} 
where ${\cal A} (\rho^{0} (\omega))$ is the amplitude of the two mixed
Breit-Wigner distributions and $m$ is the $\pi^+ \pi^-$ invariant mass. 
In Fig.~\ref{fig1}, the  $\pi^+ \pi^-$ invariant mass spectra for
${\rho}^0 - {\omega}$ mixing is displayed.
Because of the very narrow width of the $\omega$,  
(${\Gamma}_{\omega} = 8.44 \ {\rm MeV} $), we notice a high and 
narrow peak at the $\omega$ pole 
($\approx 782 \ {\rm MeV} $). 
%
\subsection{Density matrix $\boldsymbol{h_{\lambda,\lambda^{'}}}$}\label{sec5.3}
\begin{figure}
\centering\includegraphics[height=20.0cm, width = 15.0cm,clip=true]{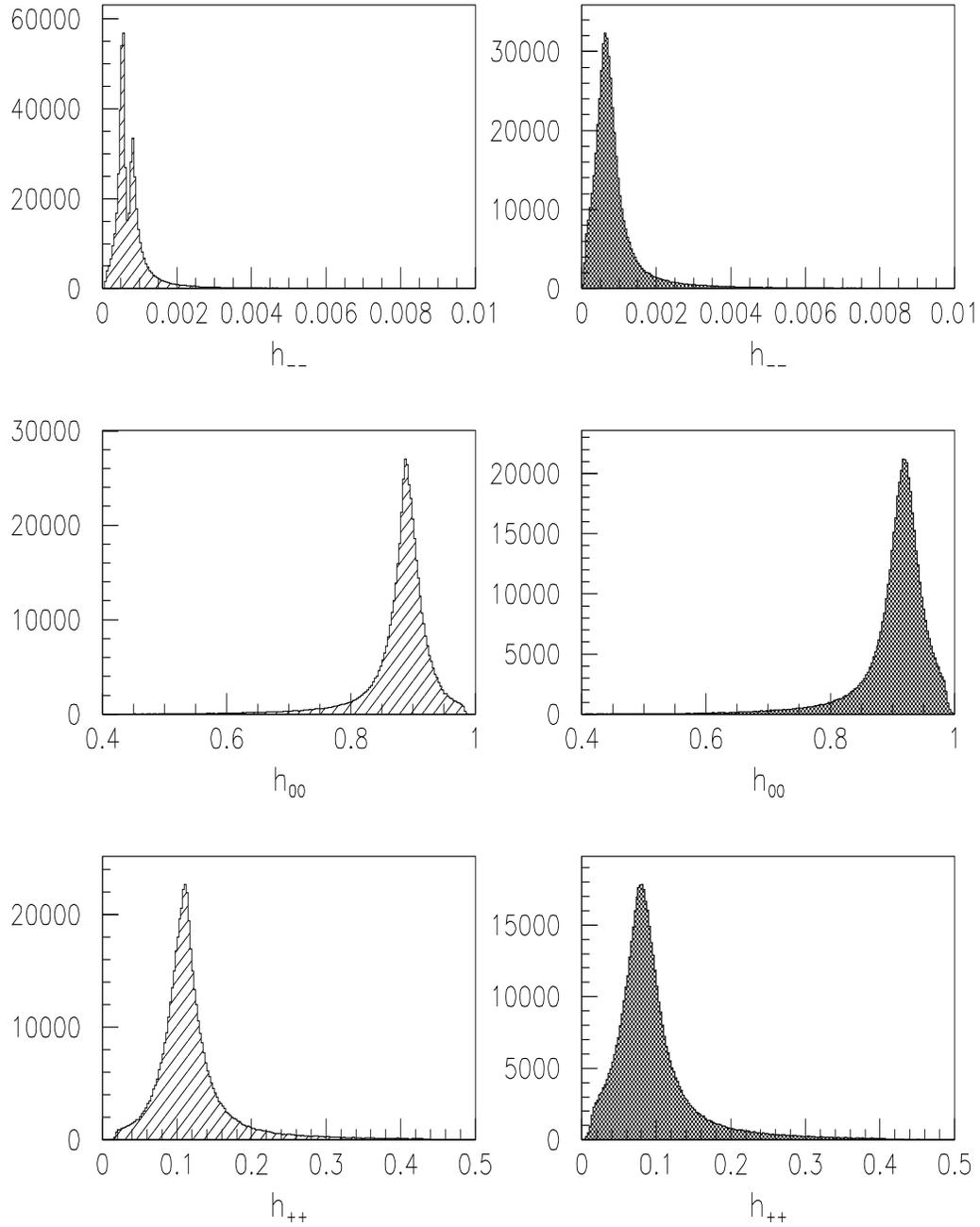}
\caption{Spectrum of $ h_{--}, h_{00}, h_{++}$. Histograms on the  left
  correspond to the channel $B^0 \rightarrow \rho^{0}(\omega) K^{* 0}$
  where the parameters used 
  are: ${q^2}/{{m^2}_b} =  0.3$, $N_c^{eff} = 2.84$, $\rho= 0.229,
  \eta=0.325$ and form factors from the GH model. 
  Histograms on the right correspond
  to the channel $B^+ \rightarrow \rho^{0}(\omega) \rho^+$ for the same
  parameters with $N_c^{eff} = 2.01$.}
\label{fig2}
\end{figure}

%
Three main parameters remain free in our simulations: the ratio ${q^2}/{m^2_b}$
(related to the penguin diagrams), the form factor model (GH or BSW) and
the effective number of colors, $N_c^{eff}$ 
(associated with the factorization hypothesis). 
The histograms plotted in Fig.~\ref{fig2}  display spectra of the diagonal
and normalized density matrix elements $h_{i,i}$, for the channels $B^0
\to \rho^0 (\omega) K^{* 0}$ (left hand-side) and $B^+ \to \rho^0 (\omega)
\rho^{+}$ (right hand-side).   
The input numerical parameters are  ${q^2}/{{m^2}_b} =  0.3$, $N_c^{eff} = 2.84$
(left hand-side figure) or  $N_c^{eff} = 2.01$ (right hand-side figure), and the
GH form factor model is applied for both decays. Note also that the average
values of CKM parameters $\rho$ and $\eta$ are used.
The wide spectrum of values of the density matrix element 
$h_{{\lambda}, {\lambda}}$, 
is caused by the resonance widths (especially that of the $\rho$)  
which provides, in turn, 
a large spectrum for the common momentum $p_V$ in the $B$ rest-frame. 
Whatever the $\rho^{0}(\omega) V_2$ channel is, $h_{00} = {|H_{0}|}^2$, 
which corresponds to longitudinal polarization, is the dominant value.
Numerically, for the $B^0 \to \rho^0 (\omega) K^{* 0}$ decay, the mean
value of $h_{00}$ is around $0.87$ while it is of order  
$0.90$ for $B^+ \to \rho^0 (\omega) \rho^{+}$.
The dominance of the longitudinal polarization has 
already been confirmed experimentally,   
since recent experimental data related to the channel 
$B \rightarrow J/{\psi} K^*$ show clearly that 
the longitudinal decay amplitude dominates in that case,
with ${|H_0|}^2 = 0.59 \pm 0.06 \pm 0.01$~\cite{CDF}.
Extrapolating these results to the charmless vector 
meson final states requires 
some modifications of the form factors 
without a big change of the relative contributions of 
the polarization states. 
Regarding $h_{--} = {|H_{-1}|}^2$, it represents less than $0.5 \%$ of the
  total amplitude for both decays.  This numerical result is 
confirmed by complete analytical calculations.

\begin{figure}
\centering\includegraphics[height=20.0cm, width = 15.0cm,clip=true]{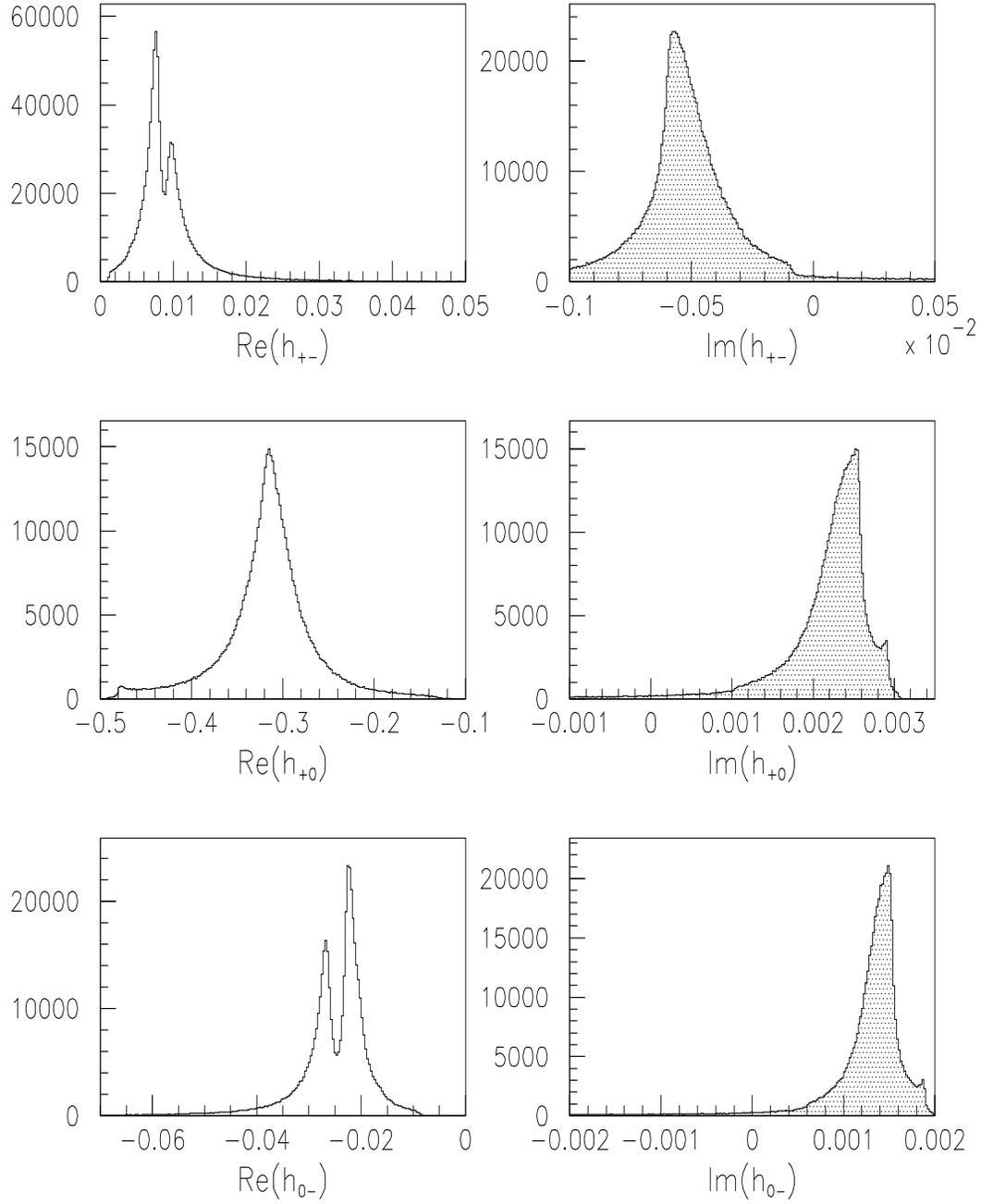}
\caption{Spectrum of $ \mathscr{R}\!e{(h_{ij})} \;  {\rm  and} \;  \mathscr{I}\!m{(h_{ij})}$
  where $i \neq j$.   Histograms 
  correspond to channel $B^0 \rightarrow \rho^{0}(\omega) K^{* 0}$ where
  the used 
  parameters are: ${q^2}/{{m^2}_b} =  0.3$, $N_c^{eff} = 2.84$, $\rho=0.229,
  \eta=0.325$ and form factors from the GH model. }
\label{fig3}
\end{figure}

\begin{figure}
\centering\includegraphics[height=20.0cm, width = 15.0cm,clip=true]{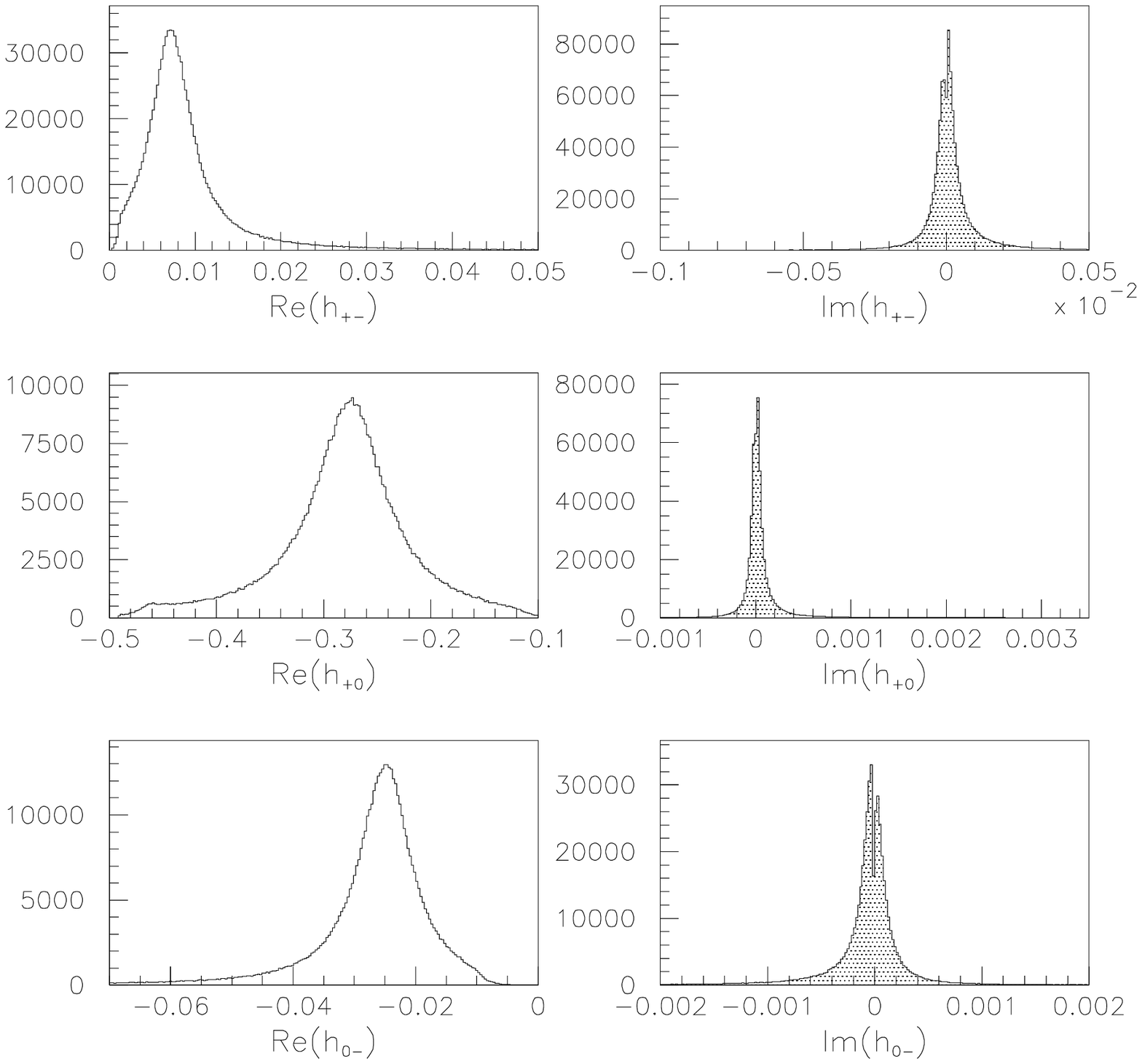}
\caption{Spectrum of $ \mathscr{R}\!e{(h_{ij})} \;  {\rm  and} \;  \mathscr{I}\!m{(h_{ij})}$
  where $i \neq j$.   Histograms 
  correspond to the channel $B^+ \rightarrow \rho^{0}(\omega) \rho^+$ where
  the used
  parameters  are: ${q^2}/{{m^2}_b} =  0.3$, $N_c^{eff} = 2.01$, $\rho=0.229,
  \eta=0.325$ and form factors from the GH model.}
\label{fig4}
\end{figure}

In  Figs.~\ref{fig3} and~\ref{fig4}, the real and imaginary parts of the
non-diagonal and normalized  density matrix elements
$h_{i,j}$ are shown for the channels $B^0 \to \rho^0 (\omega) K^{* 0}$  
and $B^+ \to \rho^0 (\omega)
\rho^{+}$, respectively. 
The input parameters are the same as previously mentioned.   
The main feature of the non-diagonal matrix elements, $h_{i,j}$, is the
smallness of both the imaginary and real parts -- 
the imaginary part being at least
one order of magnitude smaller than the real part one.
{}For the $B^+ \to \rho^0 (\omega)\rho^{+}$ decay, we observe that the mean
value of all the imaginary parts is zero, whereas it can vary for the other
decay. Note also that each of the three real parts are quite similar
{}for both decays. 
Because of the tiny value of $h_{--} = {|H_{-1}|}^2$, 
the moduli of the non-diagonal elements,   
$h_{+-} = {H_+}{H^{*}_{-}}$ and $h_{0-} = {H_0}{H^{*}_{-}}$, are very small,
while the modulus of $h_{+0} = H_{+}{H^{*}_{0}}$ is around $0.3$ for 
both decays.   
As a first conclusion, 
the general behavior of the density matrix seems to be similar
whatever the decay is. Experimentally, only the mean values of the diagonal
elements and $h_{+-}$ will be able to be measured through 
the angular distributions.

\begin{figure}
\centering\includegraphics[height=18.0cm, width = 14.0cm,clip=true]{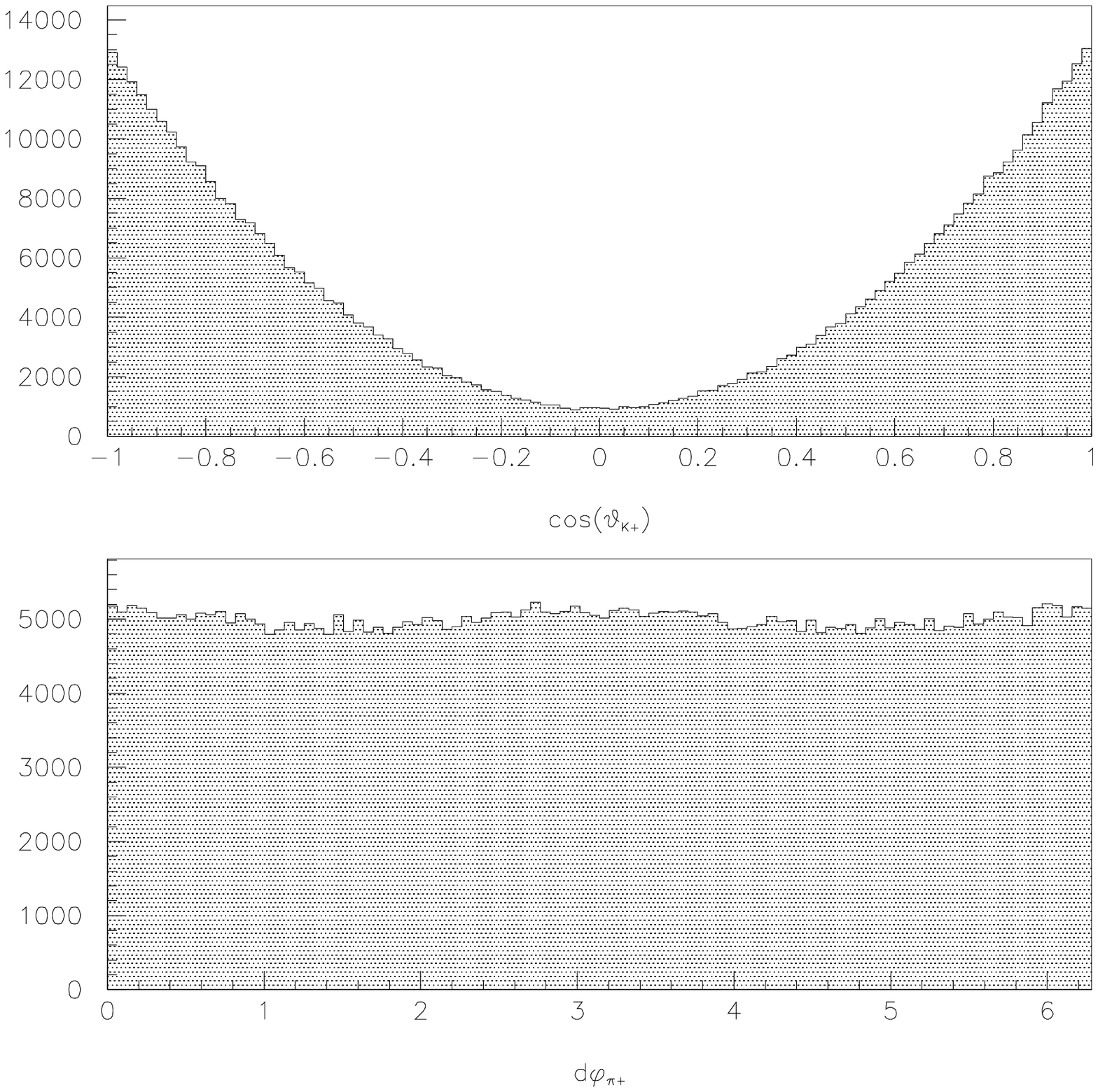}
\caption{Spectrum of polar angle (upper figure) and azimuthal angle (lower
  one) in the  helicity frame for the channel $B^0 \rightarrow \rho^{0}(\omega)
  K^{* 0}$. Parameters used are: ${q^2}/{{m^2}_b} =  0.3$, $N_c^{eff} = 2.84$, $\rho=0.229,
  \eta=0.325$ and form factors from the GH model.}
\label{fig5}
\end{figure}

\begin{figure}
\centering\includegraphics[height=18.0cm, width = 14.0cm,clip=true]{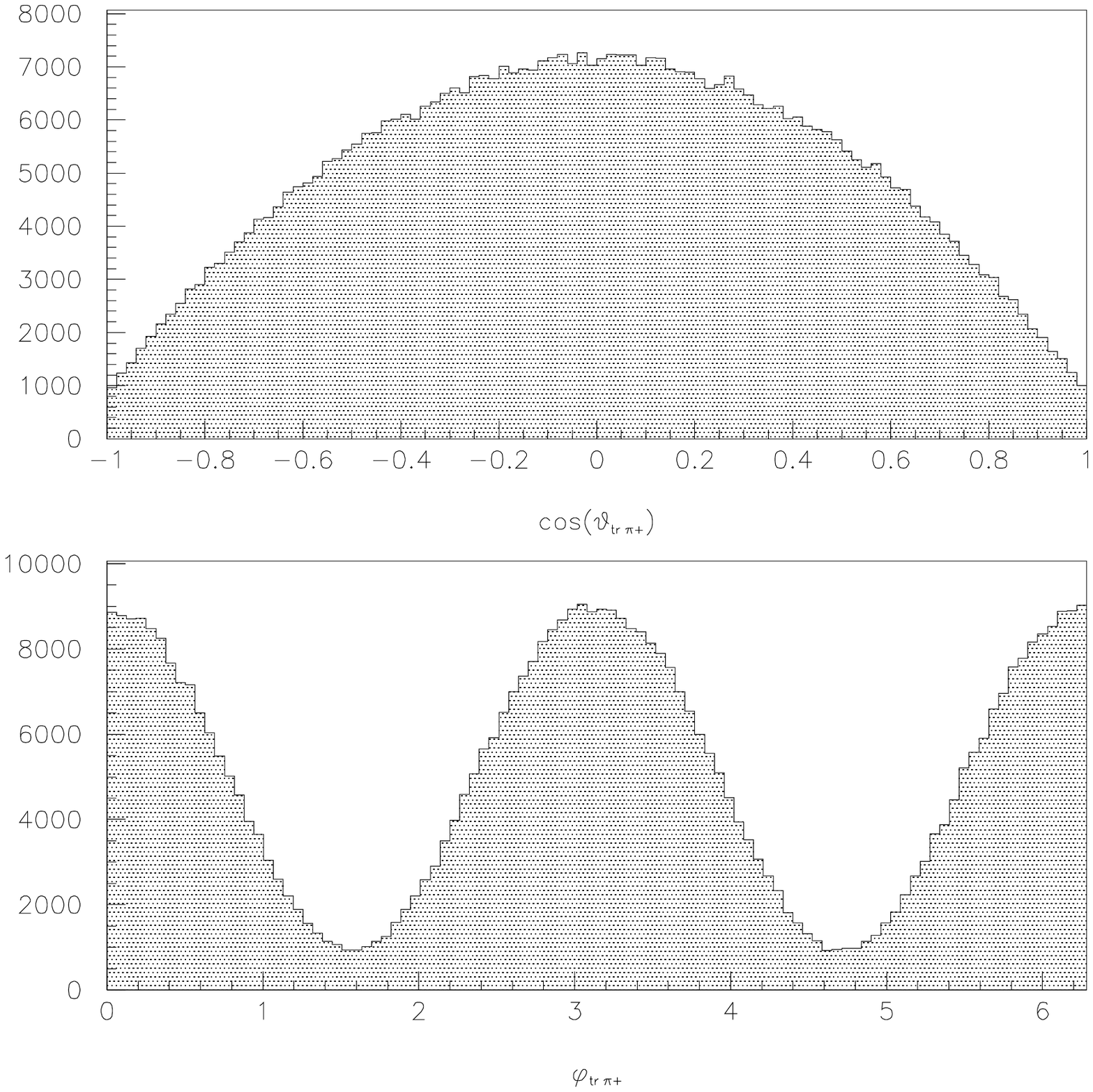}
\caption{Spectrum of polar angle (upper figure) and azimuthal angle (lower one) in the  transversity
frame for the channel $B^0 \rightarrow \rho^{0}(\omega)
  K^{* 0}$. Parameters  are: ${q^2}/{{m^2}_b} =  0.3$, $N_c^{eff} = 2.84$, $\rho=0.229,
  \eta=0.325$ and form factors from the GH model. }
\label{fig6}
\end{figure}

These angular distributions are plotted in Figs.~\ref{fig5} and~\ref{fig6} in
the helicity frame and in the transversity frame, respectively
for $B^0 \to \rho^0 (\omega) K^{* 0}$ and for the usual input parameters.
Their normalized pdfs have been displayed above in Eq.~(\ref{eq10}). 
As a consequence of the small value of $\langle h_{+-} \rangle$,  
the azimuthal angle distribution in the helicity frame is nearly flat, whereas
it is sinusoidal in the transversity frame.
{}From the distribution as a function of 
polar angle (in the TF) displayed in
Eq.~(\ref{eq13}), one can infer a mean value of the $H_{T}$ amplitude.  
This represents
an additional piece of information through which one can access
the dynamics of $B(\bar{B})$ decays into
two charmless hadrons.
%
 \section{Branching ratio and asymmetry in $\boldsymbol{B}$ decays into two vector mesons}\label{sec6}
%
The analytic expressions for the density matrix elements, $h_{ij}$ allow
us to calculate the hadronic branching ratios $\mathscr{B}(B \to
\rho^{0}(\omega) V_{2})$ and to estimate the asymmetries related to $B$ 
and $\bar{B}$ decays.
All these physical observables depend primarily on a subset of the parameters
mentioned previously, such as  
the form factors, the ratio ${q^2}/{{m_b}^2}$ 
(where $q^2$ is the mass of the virtual gluon
in the penguin diagram), the effective number of colors, $N_c^{eff}$ 
(used as a free parameter in the framework of the factorization hypothesis),
and the CKM matrix element parameters $\rho$ and $\eta$.

\subsection{Branching ratio: results and discussions}\label{sec6.1}
%

Departing from the definition of the branching ratio ($\mathscr{B}{(B \rightarrow f)}$),
\begin{equation}\label{eqO1}
\mathscr{B} {(B \rightarrow f)}  =  \frac{\Gamma{(B \rightarrow f)}}{\Gamma{(B
    \rightarrow All)}}\ ,
\end{equation}
the width ${\Gamma{(B \rightarrow V_1 V_2)}}$  
can be inferred from its differential form given by the
standard relation~\cite{Jackson}:
\begin{eqnarray}\label{eqO2}
 d{\Gamma}(B \rightarrow V_1 V_2) \ = \  {\frac{1}{8{\pi}^2 M}} \ {|{\cal M}(B \rightarrow V_1 V_2)|}^2 \ {\frac{d^3{\vec p_1}}{2E_1}} \ {\frac{d^3{\vec p_2}}{2E_2}}
 \ {\delta}^4{(P- p_1 - p_2)}\ .
\end{eqnarray}
In Eq.~(\ref{eqO2}),  $P = (M,{\vec 0})$, 
where $M = m_b$ and ${(E_1, {\vec p_1})}$ and ${(E_2, {\vec p_2})}$
are the 4-momenta of $V_1$ and $V_2$, respectively,
in the $B$ rest frame. Because of the large width of the
${\rho}^0$ meson ($\Gamma_{\rho} \approx 150$ MeV) 
and the $K^{*}$ meson  
($\Gamma_{K^{*}} \approx 50$ MeV), 
the energy, $E_i$, and the momentum, $p_i$, of each vector meson vary
according to the generated event. Computation of $\Gamma{(B \rightarrow
 {\rho}^0{(\omega)}V_2)}$ could not be done analytically but numerically 
by Monte-Carlo methods. A total number of $50000$ 
events have been generated in order to obtain a precise estimate
of this decay width.

In Tables~\ref{tab5} and~\ref{tab6} we list (respectively) the branching
ratios for $\bar B  \rightarrow
{\rho}^0(\omega) {\bar V_2}$ and  $B \rightarrow {\rho}^0(\omega) V_2$
and their dependence on 
the form factor models (BSW and GH), $q^2/m_b^2$, $N_c^{eff}$ and
the average values of the CKM parameters $\rho$ and $\eta$.  
{}For a fixed value of ${q^2}/{{m_b}^2}$, there are important 
variations of the branching ratios, depending on 
the form-factor model. They can vary by up to a factor two.
In the framework of a given form-factor model, some branching ratio  
modifications appear with ${q^2}/{{m_b}^2}$,
especially in the channels including a $K^*$. However, these changes
do not exceed $34\%$. Regarding the ratio between $\mathscr{B}(B^0 \rightarrow
{\rho}^0(\omega) K^{* 0})$ and $\mathscr{B}(B^+ \rightarrow {\rho}^0(\omega)
K^{* +})$, its value is found to be of the order $0.40$ for the BSW model
and $0.34$ for the GH model.

Finally, we observe that the relative difference between two conjugate branching ratios, 
$\mathscr{B}{(B \rightarrow f)}$ and  
$\mathscr{B}{({\bar B} \rightarrow {\bar f})}$, is almost independent of the form-factor
models, for a fixed value of ${q^2}/{{m_b}^2}$. 
It can be computed from the two
tables just mentioned and, usually, 
it does not exceed $20\%$. The exception is for the 
$K^{*\pm} {\rho}^0{(\omega)}$ channels, where it reaches $39\%$.

\subsection{Asymmetry: results and discussions}\label{sec6.2}
%

 A search for direct $CP$ violation requires asymmetries between 
conjugate final states coming from $B$ and 
$\bar B$ decays respectively. In our case, 
these searches are performed in two complementary ways. 
We consider first the global $CP$-violating asymmetry $a_{CP}$, calculated from
branching ratios:
\begin{eqnarray}\label{eqO3a}          
 a_{CP} =   \frac{\mathscr{B}(B \rightarrow f) - \overline{\mathscr{B}}(\bar B
     \rightarrow {\bar f})}{\mathscr{B}(B \rightarrow f) +
 \overline{\mathscr{B}} (\bar B \rightarrow {\bar f})}\ . 
\end{eqnarray}
Secondly, we use the partial widths of $B {(\rightarrow f)}$ 
and $\bar B {(\rightarrow {\bar f})}$,  
calculated as described above together with 
the differential asymmetries investigated 
as a function of the ${\pi}^+ {\pi}^-$ invariant mass 
in the whole range of the ${\rho}^0$ Breit-Wigner resonance. Hence,
$a_{CP}(m)$ takes the following form:
\begin{eqnarray}\label{eqO3b}          
 a_{CP}(m) =   \frac{{\Gamma}_m{(B \rightarrow f)} - {\bar \Gamma}_m{(\bar B
     \rightarrow {\bar f})}}{{\Gamma}_m{(B \rightarrow f)} +
 {\bar \Gamma}_m{(\bar B \rightarrow {\bar f})}}\ , 
\end{eqnarray}    
where $m$ is the ${\pi}^+ {\pi}^-$ invariant mass.  ${\Gamma}_m{(B \rightarrow
  f)}$  and ${\bar \Gamma}_m{(\bar B \rightarrow {\bar f})}$ in
Eq.~(\ref{eqO3b}) are the  partial widths written as a function of $m$.

\begin{figure}
\centering\includegraphics[height=16.0cm, width = 14.0cm,clip=true]{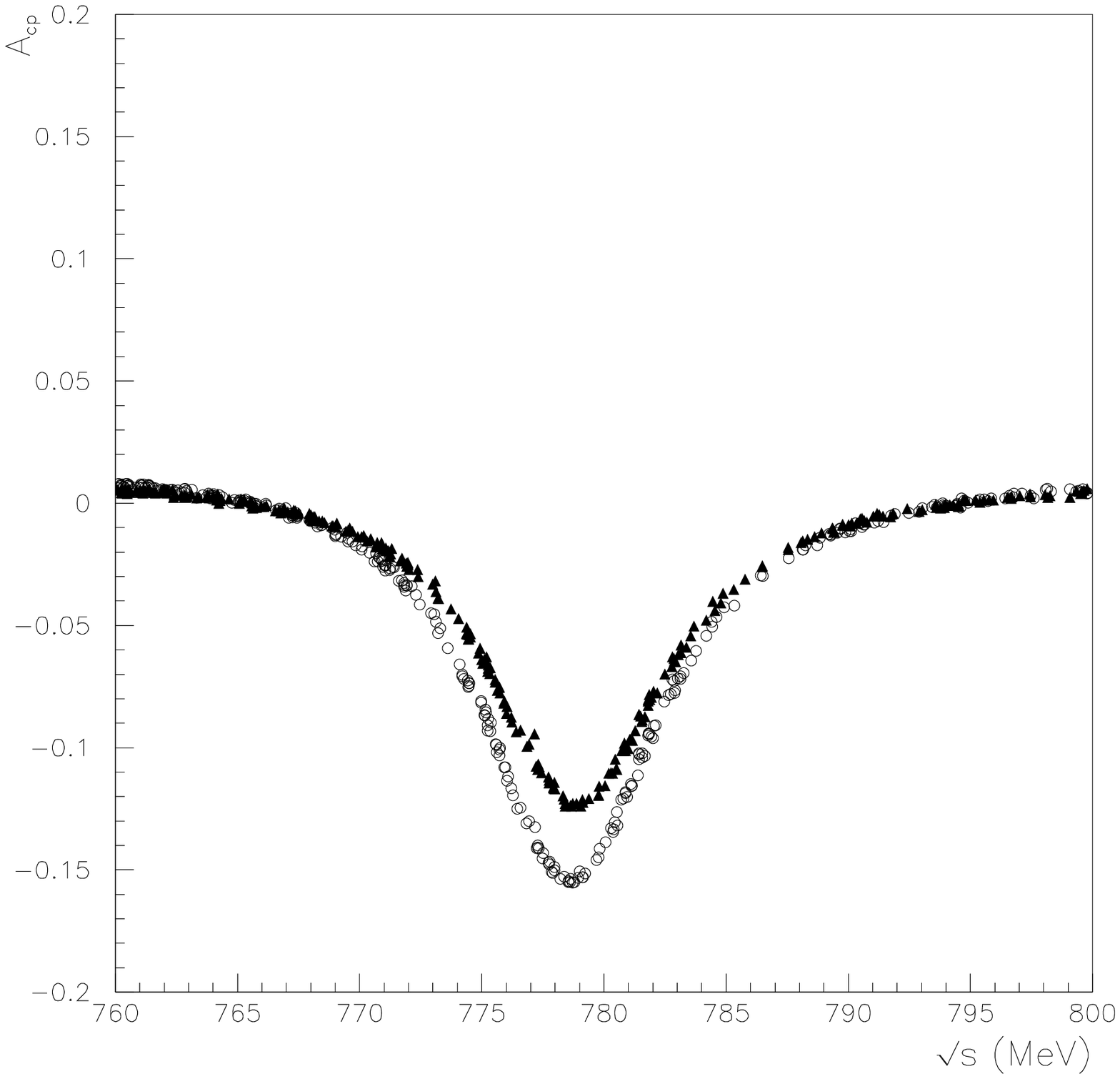}
\caption{ $CP$-violating asymmetry parameter $a_{CP}(m)$, as a function of the
  ${\pi}^+ {\pi}^-$
invariant mass in the vicinity of the $\omega$ mass region for the channel $B^0
\rightarrow \rho^{0}(\omega)
  K^{* 0}$. Parameters  are: ${q^2}/{{m^2}_b} =  0.3$, $N_c^{eff} = 2.84$, $\rho= 0.229,
  \eta=0.325$. Solid triangles up and circles correspond to the BSW and GH form
  factor models respectively.}
\label{fig7}
\end{figure}

\begin{figure}
\centering\includegraphics[height=16.0cm, width = 14.0cm,clip=true]{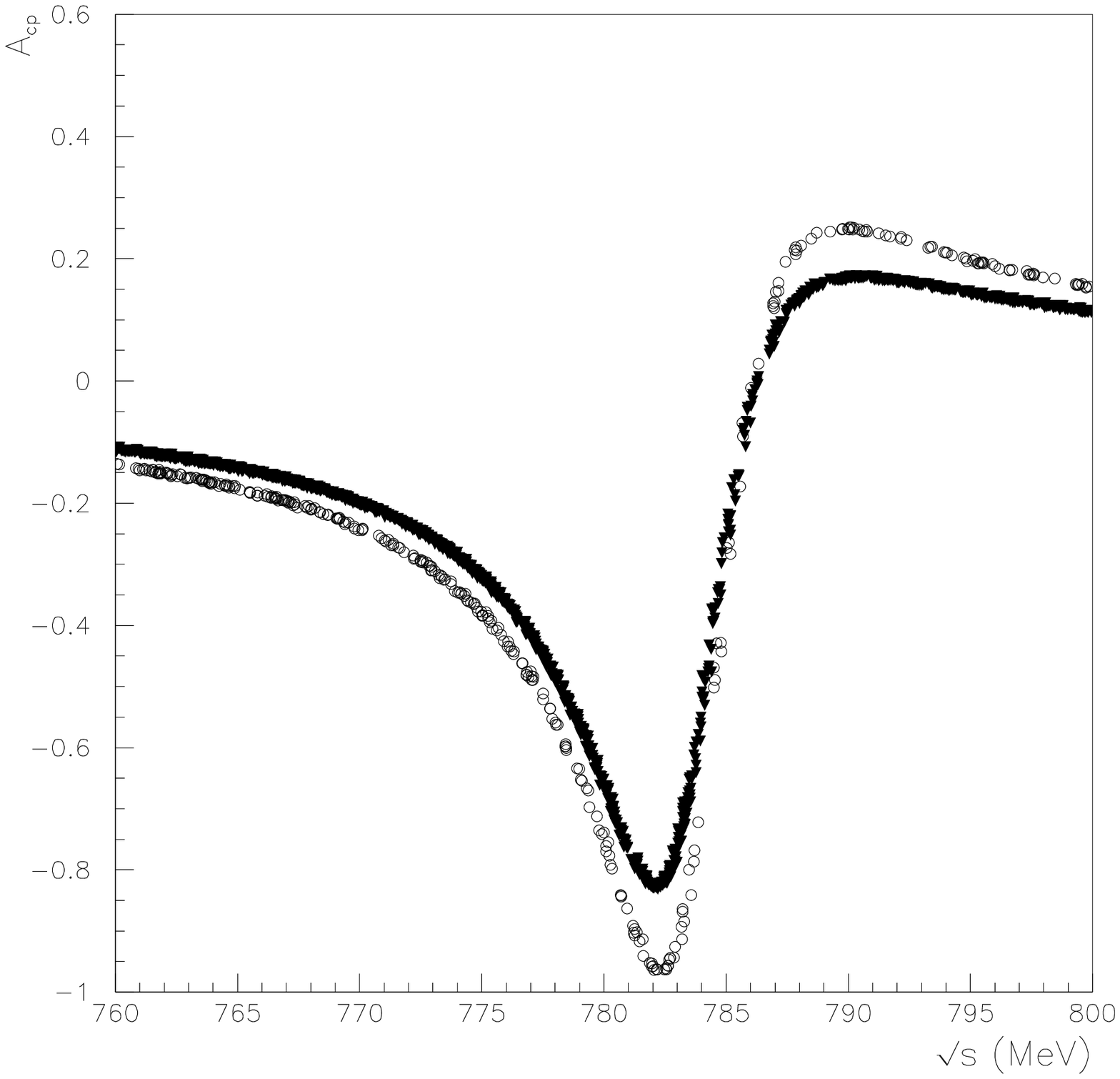}
\caption{ $CP$-violating asymmetry parameter $a_{CP}(m)$, as a function of the
 ${\pi}^+ {\pi}^-$
invariant mass in the vicinity of the $\omega$ mass region for the channel $B^+ \rightarrow \rho^{0}(\omega)
  \rho^+$. Parameters  are: ${q^2}/{{m^2}_b} =  0.3$, $N_c^{eff} = 2.01$,
  $\rho= 0.229,  \eta=0.325$. Solid triangles down and circles 
  correspond to the BSW and GH form
  factor models respectively. }
\label{fig8}
\end{figure}

In Table~\ref{tab7} we list the global $CP$-violating  
asymmetry between the $B$ and $\bar B$ decays 
for the channels under investigation. 
It can be noticed that, for a fixed value of ${q^2}/{{m_b}^2}$, the two form
factor models provide quite similar results. For 
different ${q^2}/{{m_b}^2}$ values, the corresponding
results could vary, especially in the $K^{*\pm} {\rho}^0{(\omega)}$ channels.
In Figs.~\ref{fig7} and~\ref{fig8} we show, respectively, the histogram of the
direct $CP$-violating asymmetry parameter $a_{CP}(m)$, for the decays $B^0
 \rightarrow {\rho}^0(\omega) K^{* 0}$ and $B^+ \rightarrow
 {\rho}^{0}(\omega) \rho^{+}$, as a function of the 
${\pi}^+ {\pi}^-$ invariant mass in the $\omega$ mass region and for both form
{}factor models. The asymmetry reaches its maximum when $\sqrt s$ is around
$780$ MeV. However, outside the displayed windows, 
the asymmetry goes to zero in any
case. The peak of the asymmetry is emphasized when the GH form factor model is
used in our simulations. For the $K^{* 0} {\rho}^0{(\omega)}$ channels,
the maximum of the $CP$ violating asymmetry is around $13\%$ 
and $16\%$, for the BSW model and
the GH model, respectively. Finally, we emphasise that
the ${\rho}^{\pm} {\rho}^0{(\omega)}$ channels present the 
most intriguing results because,
in any case, their asymmetry  is at least $80\%$ (BSW model) and can reach
$95\%$ (GH model). 
This last channel is highly recommended for a direct search for $CP$ violation.

\section{Perspectives and conclusions}\label{sec7}

We have studied direct $CP$ violation in decay process such as $B \to
  \rho^0(\omega) V_2 \to \pi^+ \pi^- V_2$,
where $V_2$ is either $K^{* 0,\pm}$ or $\rho^{\pm}$,  
with the inclusion of  ${\rho}^{0}-{\omega}$ mixing. When the invariant mass
of the $\pi^{+}\pi^{-}$ pair is in the vicinity of the $\omega$ resonance, it
is found that the  $CP$-violating asymmetry, $a_{CP}(m)$, reaches its
maximum value. 
In our analysis we have also investigated the branching ratios  for the same
channels. 
Thanks to the standard helicity and transversity formalisms,   
rigorous and detailed calculations
of the $B^{0 {\pm}}$ decays into two charmless vector mesons have been
carried out completely. 
Using the effective Hamiltonian based on the operator product expansion
with the appropriate Wilson coefficients, 
we derived in detail the amplitudes corresponding to $B \to
 \rho^0(\omega) V_2 \to \pi^+ \pi^- V_2$ decay and 
the density matrix, $h_{\lambda \lambda^{\prime}}$ as well. 

In order to apply our
{}formalism, we used a Monte-Carlo method for all the  numerical simulations.
Moreover, we dealt at length with the  
uncertainties coming from the input parameters. In particular, 
these include the Cabibbo-Kobayashi-Maskawa matrix
element parameters, $\rho$ and $\eta$, the effective number of colors,  
$N_c^{eff}$, coming from the naive factorization and  two  phenomenological
models in order to show the possible dependence on form factors, GH or BSW.
These form factors vary slightly according to the final states. Recall that this work was achieved by 
applying a phenomenological treatment, where some assumptions regarding the evaluation 
of the hadronic matrix elements have been made. In this approach,  
 corrections associated with the limit of validity of the
factorization hypothesis were parameterized phenomenologically and may involve large
uncertainties.

As a major result, the predominance of the
longitudinal polarization,  $h_{00}$, has  been pointed out in  all the 
investigated decays. We also found a large direct $CP$-violating
asymmetry in these $B$ decays into two charmless vector mesons.  
We stress that,  
without the inclusion of ${\rho}^{0}-{\omega}$ mixing, we would not have a
large $CP$-violating asymmetry. Finally, we
predicted branching ratios to be of the order $0.7-2.1 \times 10^{-6}$ for
$K^{* 0} \rho^{0}(\omega)$ and of the order $2.3-6.6 \times 10^{-6}$ for
$K^{* \pm} \rho^{0}(\omega)$  (depending on the different
phenomenological models).  For the channel $\rho^{\pm} \rho^{0}(\omega)$, we
{}found the branching ratios to  be of the order $11-24 \times 10^{-6}$.

Two main conclusions can be drawn. The first is the relative
importance of the form factor  model which is
used, since some branching ratios in $B \to \rho^{0}(\omega) V_2$ could change
by up to a factor two. The second is the important role
of  ${\rho}^{0}-{\omega}$ mixing, which can enhance 
considerably the asymmetry parameter ${a_{CP}}$, 
between the conjugate final states coming, respectively, from 
$B$ and $\bar B$ decays.

Beside the ``standard'' ways to look for direct $CP$ violation, 
such as the difference between branching ratios and/or
discrepancies in the angular distributions of the decay products, 
we have presented a detailed discussion of a new method.
This involves the variation of
${a_{CP}}$ as a function of the ${\pi}^+ {\pi}^-$ invariant 
mass over the whole range of the ${\rho}^0$ 
resonance~\cite{Gardner:1997yx,refolive}. 
We believe that this method will be very
{}fruitful for future experiments and has already been 
implemented in the generator of the LHCb experiment.
Indeed, we look forward to being able
to apply the formalism developed here to the analysis of 
experimental data for decays such as $B \to \rho^0 (\omega) V_2$ (with
$V_2$ being either a $K^{*}$ or a $\rho^{\pm}$) in 
the near future.

\subsubsection*{Acknowledgements}
This work was supported in part by the Australian Research Council and the
University of Adelaide. The LHCb Clermont-Ferrand group would like to acknowledge G. Menessier from  the
LPTM, for many illuminating discussions regarding the exciting question of FSI
in hadronic physics.
%
\section*{Appendix}\label{sec8}
%
\appendix
%
\section{Practical calculations of the helicity amplitudes}\label{seca}
%
The helicity formalism in the case of vector mesons 
requires the introduction of 
three polarization four-vectors for each spin 1 particle~\cite{DEWITT}:
\begin{equation}\label{aeq1}
{\epsilon}(1) = (0, \vec {\epsilon}(1))\ , \ \ {\epsilon}(2) = (0, \vec
{\epsilon}(2))\ , \; {\rm and} \;\; {\epsilon}(3)
={\left(|{\vec k}|/m , E {\hat {k}}/m \right)}\ .
\end{equation}
They also satisfy the following relations as well:
\begin{equation}\label{aeq2}
 {{\epsilon}(i)}^2 = -1\ ,  \; {\rm and} \;\; {\epsilon}(i) \cdot {\epsilon}(j) = 0\ , \ \
 \mathrm{with} \  i \neq j\ ,
\end{equation}
where $m, E$ and $\vec k$ are respectively the mass, the energy and the
momentum of the vector meson. $\hat {k}$ is
defined as the unit vector along the vector momentum, $\hat {k} = {\vec k}/{|\vec k|}$.
The three vectors $\vec {\epsilon}(1), \vec {\epsilon}(2)$ and $\vec
{\epsilon}(3) = {E {\hat {k}}}/m$ 
form an orthogonal basis.  
$\vec {\epsilon}(1)$ and $\vec {\epsilon}(2)$ are the {\it transverse polarization} vectors while  
$\vec {\epsilon}(3)$ is the {\it longitudinal polarization} vector. 
These three vectors allow one to define the {\it helicity basis}:
\begin{equation}\label{aeq3}
\epsilon(+) = \frac{\left({\epsilon}(1) + i \; {\epsilon}(2) \right)}{\sqrt 2}\ , \ \ {\epsilon}(-) =
\frac{\left({\epsilon}(1) - i \; {\epsilon}(2) \right)}{\sqrt 2}\ , \ \ {\rm and}
\ \ 
{\epsilon}(0) = {\epsilon}(3)\ .
\end{equation}
These 4-vectors are {\it eigenvectors} of the helicity operator $\mathcal {H}$ 
corresponding, respectively, to the eigenvalues  $ \lambda = +1, -1$ and $0$.
In the $B^{0 {\pm}}$ rest-frame, the vector mesons have opposite momentum 
$ \vec {k}_1   =   -\vec {k}_2$ 
and their respective polarization vectors are {\it correlated}. This implies
the following expressions,
$\;\;\;\;$ 
$$\vec{k}_{K} = -\vec{k}_{\rho} = \vec{k} = \left(
                               \begin{array}{c}
                                 k\sin\theta\cos\phi\\
                                 k\sin\theta\sin\phi\\
                                 k\cos\theta
                                \end{array}
                               \right)\ , $$

where $\theta$ and $\phi$ are respectively polar and azimuthal angles of the
produced $K^{*}$. In our case, one has for the transversal polarization
vectors ($K^{*}$ and $\rho$) the expressions:
$$\vec{\epsilon}_{K}(1) = \left(
                               \begin{array}{c}
                                 \cos\theta\cos\phi\\ 
                                 \cos\theta\sin\phi\\ 
                                 -\sin\theta 
                                \end{array} 
                               \right) = \vec{\epsilon}_{\rho}(1)\ ,$$

\noindent and,

$$\vec{\epsilon}_{K}(2) = \left(
                               \begin{array}{c} 
                                 -\sin\phi\\ 
                                 \cos\phi\\ 
                                  0 
                                \end{array} 
                               \right) = -\vec{\epsilon}_{\rho}(2)\ .$$
\noindent Regarding the longitudinal polarization, ${\epsilon}_{K}(3)$ and
${\epsilon}_{\rho}(3)$ take the form:
\begin{equation}\label{aeq4}
\epsilon_{K}(3) =
\left(\frac{|\vec{k}|}{m_{K}},\frac{E_K}{m_K}\hat{k}\right)\ , 
\ \ \ \epsilon_{\rho}(3) =
\left(\frac{|\vec{k}|}{m_{\rho}},\frac{E_{\rho}}{m_{\rho}}(-\hat{k})\right)\ .
\end{equation}
By applying the relations from Eq.~(\ref{aeq3}), one can expressed 
vectors $\vec{\epsilon}(i)$ in the 
helicity basis and one gets $\vec{\epsilon}(\pm)$: \\
\vspace{-0.5cm}
\begin{center}
\begin{equation}\label{aeq5}
\vec{\epsilon}_{K}(+) = \left(
                               \begin{array}{c} 
                                 \cos\theta\cos\phi-i\sin\phi\\ 
                                 \cos\theta\sin\phi+i\cos\phi\\ 
                                  -\sin\theta 
                                \end{array} 
                               \right) / \sqrt{2} =
                               \vec{\epsilon}_{K}^{\ *}(-) = 
                                \vec{\epsilon}_{\rho}(-)\ ,
\end{equation}
\end{center}
\begin{center}
\begin{equation}\label{aeq6}
\vec{\epsilon}_{K}(-) = \left(
                               \begin{array}{c} 
                                 \cos\theta\cos\phi+i\sin\phi\\ 
                                 \cos\theta\sin\phi-i\cos\phi\\ 
                                  -\sin\theta 
                                \end{array} 
                               \right) / \sqrt{2} = \vec{\epsilon}_{K}^{\
                                 *}(+) = 
\vec{\epsilon}_{\rho}(+)\ .
\end{equation}
\end{center}
The weak hadronic amplitude is therefore decomposed, in 
the helicity basis, according to the general method
developed by Bauer, Stech and Wirbel~\cite{ref20}. This 
will allow one to obtain two interesting results. Firstly, one can isolate 
the contribution of each helicity state to the total amplitude. Secondly, 
the contributions of the {\it tree} and {\it penguin} operators 
to the total amplitude can be separated via 
the helicity states.

The knowledge of the main input parameters $\rho, \eta, A ,
{\sin{\theta}_c}(=\lambda_c) $ 
and the masses and 
widths of the intermediate resonances allow a complete determination of the three helicity amplitudes 
$H_{\lambda}\bigl(B \rightarrow \rho^{0}(\omega) V_2 \bigr)$,
where the helicity $\lambda$ can take the values -1,0 or +1. 
%
\section{Channel $\boldsymbol{B^{\pm} \rightarrow \rho^{0}(\omega) \rho^{\pm}}$}\label{secb}
%
The formalism applied in case of $B  \rightarrow \rho^{0}(\omega) K^{*}$
can be extend to  $B^{\pm} \rightarrow \rho^{0}(\omega)
\rho^{\pm}$. Nevertheless, in the last case one has  $b \rightarrow d$
transition instead of  $b \rightarrow s$. 
The amplitude 
$H_{\lambda}\bigl(B \rightarrow \rho^{0}(\omega) V_2 \bigr)$ 
has the form:
\begin{multline}\label{aeq7}
H_{\lambda=^0_{\pm{1}}} \bigl(B \rightarrow \rho^{0}(\omega) V_2 \bigr)= \\
 A\lambda^{3} \Bigg \lbrace \bigg \lbrack 
R_1^{\rho} B^{\rho}_{\lambda=^0_{\pm{1}}}+ R_2^{\rho}C^{\rho}_{\lambda=^0_{\pm{1}}}\bigg \rbrack 
+ \ i\ \bigg \lbrack I_1^{\rho} B^{\rho}_{\lambda=^0_{\pm{1}}} + I_2^{\rho}
C^{\rho}_{\lambda=^0_{\pm{1}}} \bigg \rbrack \Bigg \rbrace \hspace{2.7cm} \\
+ \frac{
\tilde{\Pi}_{\rho
 \omega}}{(s_{\rho}-m_{\omega}^{2})+im_{\omega}\Gamma_{\omega}}
\Biggl[
A\lambda^{3} \Bigg \lbrace \bigg \lbrack R_1^{\omega}
B^{\omega}_{\lambda=^0_{\pm{1}}}  + R_2^{\omega} C^{\omega}_{\lambda=^0_{\pm{1}}}\bigg \rbrack \\
+ \ i\ \bigg \lbrack I_1^{\omega}  B^{\omega}_{\lambda=^0_{\pm{1}}} + I_2^{\omega}
C^{\omega}_{\lambda=^0_{\pm{1}}} \bigg \rbrack \Bigg \rbrace
\Biggr]\ ,
\end{multline}
where one defines,
\begin{align}\label{aeq8}
R_i^{V_1} & = (1-\frac{\lambda^2}{2})\eta  c_{t_{i}}^{V_1} +
\eta \;\mathscr{R}\!e{(c_{p_{i}}^{V_1})} - (1-\rho) \; \mathscr{I}\!m{(c_{p_{i}}^{V_1})}\ , \\
I_i^{V_1} & = (1-\frac{\lambda^2}{2})\rho  c_{t_{i}}^{V_1} +
\eta \; \mathscr{I}\!m{(c_{p_{i}}^{V_1})} + (1-\rho)\; \mathscr{R}\!e{(c_{p_{i}}^{V_1})}\ , 
\end{align}
with $V_1$ being either $\rho^{0}$ or $\omega$.
\begin{eqnarray}
{\rm If} \; \; V_1  \equiv \rho  \; {\rm and}  \; i=2 \;  {\rm then}  \; R_i^{V_1}
= I_i^{V_1}  = 0\ . 
 \end{eqnarray}
The expressions for 
$c_{t_{i}}^{V_1}$ and $c_{p_{i}}^{V_1}$, which correspond to the
investigated channel, take the following form:
\newline
\noindent for the decay $B^{-} \rightarrow  \rho^{0} \rho^{-}$: 
\begin{align}\label{aeq9}
c_{t_{1}}^{\rho} & = C^{\prime}_{1} + \frac{C^{\prime}_{2}}{N_{c}}+ 
 C^{\prime}_{2} + \frac{C^{\prime}_{1}}{N_{c}}\ , \nonumber \\
c_{p_{1}}^{\rho} & =\frac{3}{2} \Bigl(C^{\prime}_{7}+\frac{C^{\prime}_8}{N_{c}}
+C^{\prime}_{9} + \frac{C^{\prime}_{10}}{N_{c}}+
C^{\prime}_{10} + \frac{C^{\prime}_{9}}{N_{c}}\Bigl)\ .
\end{align}
\newline
\noindent In the case of $\omega$
production, one obtains the linear combinations of the effective Wilson
coefficients:
\newline
\noindent for the decay $B^{-} \rightarrow  \omega \rho^{-}$:
\begin{multline}\label{aeq10}
c_{t_{1}}^{\omega}  = C^{\prime}_{2} + \frac{C^{\prime}_{1}}{N_{c}}\ , \hspace{1.em}
 c_{p_{1}}^{\omega}  = C^{\prime}_{4} + \frac{C^{\prime}_3}{N_{c}}+
\Bigl(C^{\prime}_{10} + \frac{C^{\prime}_{9}}{N_{c}}\Bigr)\ ,  \\
c_{t_{2}}^{\omega}  = C^{\prime}_{1} + \frac{C^{\prime}_{2}}{N_{c}}\ , \hspace{1.em}
  c_{p_{2}}^{\omega}  =2\Bigl(C^{\prime}_3 +
\frac{C^{\prime}_4}{N_{c}}+ C^{\prime}_5 +
\frac{C^{\prime}_6}{N_{c}}\Bigr) \hspace{16.2em}  \\ 
   + \frac{1}{2} \Bigl(C^{\prime}_{9}+\frac{C^{\prime}_{10}}{N_{c}} 
+ C^{\prime}_{7}+\frac{C^{\prime}_{8}}{N_{c}} - C^{\prime}_{10}-
\frac{C^{\prime}_{9}}{N_{c}}\Bigr)\ .  
\end{multline}
All the terms used in the appendix have been defined in Section IV.
\newpage


\newpage
%
\section*{Figure captions} 
%
\begin{itemize}
\item{ Fig.~\ref{figOO1} Transversity frame for $B \to \rho^{0} K^{*}$.}
\item{ Fig.~\ref{figO1} Tree diagram (left), and QCD-penguin diagram (right), for $B$ decays.}
\item{ Fig.~\ref{figO2}  Electroweak-penguin diagram
    (left), and electroweak-penguin diagram with coupling between $Z,\gamma$ and
    $W$ (right), for $B$ decays.}
\item{ Fig.~\ref{fig1} Spectrum of ${\rho^{0}}- {\omega}$ mixing (in
    MeV/$c^2$), simulated by the interference of two Breit-Wigner curves.}
\item{ Fig.~\ref{fig2} Spectrum of $ h_{--}, h_{00}, h_{++}$. Histograms on the  left
  correspond to the channel $B^0 \rightarrow \rho^{0}(\omega) K^{* 0}$
  where the parameters used
  are: ${q^2}/{{m^2}_b} =  0.3$, $N_c^{eff} = 2.84$, $\rho= 0.229,
  \eta=0.325$ and form factors from the 
  GH model . Histograms on the right correspond
  to the channel $B^+ \rightarrow \rho^{0}(\omega) \rho^+$ for the same
  parameters with $N_c^{eff} = 2.01$.}
\item{ Fig.~\ref{fig3} Spectrum of $ \mathscr{R}\!e{(h_{ij})} \;  {\rm  and} \;  \mathscr{I}\!m{(h_{ij})}$
  where $i \neq j$.   Histograms 
  correspond to the channel $B^0 \rightarrow \rho^{0}(\omega) K^{* 0}$
  where the parameters used 
  are: ${q^2}/{{m^2}_b} =  0.3$, $N_c^{eff} = 2.84$, $\rho=0.229,
  \eta=0.325$ and form factors from the GH model. }
\item{ Fig.~\ref{fig4} Spectrum of $ \mathscr{R}\!e{(h_{ij})} \;  {\rm  and} \;  \mathscr{I}\!m{(h_{ij})}$
  where $i \neq j$.   Histograms 
  correspond to the channel $B^+ \rightarrow \rho^{0}(\omega) \rho^+$ where the
  used parameters are: ${q^2}/{{m^2}_b} =  0.3$, $N_c^{eff} = 2.01$, $\rho=0.229,
  \eta=0.325$ and form factors from the GH model. }
\item{ Fig.~\ref{fig5} Spectrum of polar angle (upper figure) and azimuthal angle (lower
  one) in the  helicity frame for the channel $B^0 \rightarrow \rho^{0}(\omega)
  K^{* 0}$. Parameters  are: ${q^2}/{{m^2}_b} =  0.3$, $N_c^{eff} = 2.84$, $\rho=0.229,
  \eta=0.325$ and form factors from the GH model. }
\item{ Fig.~\ref{fig6} Spectrum of polar angle (upper figure) and azimuthal angle (lower one) in the  transversity
frame for the channel $B^0 \rightarrow \rho^{0}(\omega)
  K^{* 0}$. Parameters  are: ${q^2}/{{m^2}_b} =  0.3$, $N_c^{eff} = 2.84$, $\rho=0.229,
  \eta=0.325$ and form factors from the GH model. }
\item{ Fig.~\ref{fig7}  $CP$-violating asymmetry parameter $a_{CP}(m)$, as a
    function of the  ${\pi}^+ {\pi}^-$
invariant mass in the vicinity of the $\omega$ mass region for the channel $B^0 \rightarrow \rho^{0}(\omega)
  K^{* 0}$. Parameters  are: ${q^2}/{{m^2}_b} =  0.3$, $N_c^{eff} = 2.84$, $\rho= 0.229,
  \eta=0.325$. Solid triangles up and circles correspond to the BSW  and GH form
  factor models respectively. }
\item{ Fig.~\ref{fig8}  $CP$-violating asymmetry parameter $a_{CP}(m)$, as a
    function of the  ${\pi}^+ {\pi}^-$
invariant mass in the vicinity of the $\omega$ mass region for the channel $B^+ \rightarrow \rho^{0}(\omega)
  \rho^+$. Parameters  are: ${q^2}/{{m^2}_b} =  0.3$, $N_c^{eff} = 2.01$,
  $\rho= 0.229,  \eta=0.325$. Solid triangles down and circles
  correspond to the BSW and GH form
  factor models respectively. }
\end{itemize}
\newpage
%
\section*{Table  captions}
%
%
\begin{itemize}
\item{ Table~\ref{tab1} Wilson coefficients to the next-leading order.}
\item{ Table~\ref{tab2} Effective Wilson coefficients related to the tree operators, 
 electroweak and QCD-penguin operators.}
\item{ Table~\ref{tab3} Values of the CKM unitarity triangle  for limiting
    values of the CKM matrix 
elements.}
\item{ Table~\ref{tab4} Form factor  values for  $B \rightarrow \rho$, $B \rightarrow \omega$ and $B
 \rightarrow  K$ at $q^{2}=0$.}
%
\item{ Table~\ref{tab5} $ \bar{B}^0, B^-$ branching ratios (in units of
    $10^{-6}$) using either the BSW or GH form factor models, for $q^2/m_b^2 = 0.3 (0.5)$, 
with $N_{c max}^{b \to s} = 2.84 (2.82)$, $N_{c max}^{b \to d} = 2.01 (1.95)$,
 $\rho = 0.229$ and $\eta = 0.325$.
}
\item{ Table~\ref{tab6}  $ B^0 , B^+$ branching ratios (in units of $10^{-6}$)
    using either the BSW or GH form factor models, for $q^2/m_b^2 = 0.3 (0.5)$, 
with $N_{c max}^{b \to s} = 2.84 (2.82)$, $N_{c max}^{b \to d} = 2.01 (1.95)$,
 $\rho = 0.229$ and $\eta = 0.325$.
}
%
\item{ Table~\ref{tab7} Global $CP$-violating asymmetries (in percents) 
 using either the BSW or GH form factor models, for $q^2/m_b^2 = 0.3 (0.5)$, 
         with $N_{c max}^{b \to s} = 2.84 (2.82)$, $N_{c max}^{b \to d} = 2.01 (1.95)$, $\rho = 0.229$ and
         $\eta = 0.325$.}
\end{itemize}
\clearpage

%


\clearpage
%
%
\begin{table}
\begin{center}
\begin{tabular}{cccc} 
\hline 
\hline    
\multicolumn{4}{c}{$C_{i}(\mu)$ for $\mu= 5$ GeV} \\
\hline 
\hline    
 & $C_{1} $   & $-0.3125$   \\
 &  $C_{2} $   & $+1.1502$   \\
\cline{1-4}
$C_{3} $   & $+0.0174$   & $C_{5} $  & $+0.0104$ \\
$C_{4} $   & $-0.0373$   & $C_{6} $  & $-0.0459$ \\
\hline
$C_{7} $   & $-1.050 \times 10^{-5}$   & $C_{9} $  & $-0.0101$                \\
$C_{8} $   & $+3.839 \times 10^{-4}$   & $C_{10}$  & $+1.959 \times 10^{-3}$  \\
\hline
\hline
\end{tabular}
\end{center}
\caption{Wilson coefficients to the next-leading order (see the reference in text). }
\label{tab1}
\end{table}
%
%
\begin{table}
\begin{center}
\begin{tabular}{ccc} \hline \hline    
   $C_{i}^{\prime}$          & $q^{2}/m_{b}^{2}=0.3$   & $q^{2}/m_{b}^{2}=0.5$ \\
\hline
\hline
$C_{1}^{\prime} $   &   $-0.3125$                                     &$-0.3125 $  \\
$C_{2}^{\prime} $   & $+1.1502$                                        &$+1.1502$    \\
\hline
$C_{3}^{\prime} $   & $+2.433 \times 10^{-2} + 1.543 \times 10^{-3}i$  &$+2.120 \times 10^{-2} + 2.174 
\times 10^{-3}i$\\
$C_{4}^{\prime} $   & $-5.808 \times 10^{-2} -4.628 \times 10^{-3}i$  &$-4.869 \times 10^{-2} -1.552 
\times 10^{-2}i$\\
$C_{5}^{\prime} $   & $+1.733 \times 10^{-2}+ 1.543 \times 10^{-3}i$   &$+1.420 \times 10^{-2} + 5.174 
\times 10^{-3}i$\\
$C_{6}^{\prime} $   & $-6.668 \times 10^{-2}- 4.628 \times 10^{-3}i$  &$-5.729  \times 10^{-2}- 1.552 
\times 10^{-2}i$\\
\hline
$C_{7}^{\prime} $   & $-1.435 \times 10^{-4} -2.963 \times 10^{-5}i$  &$-8.340 \times 10^{-5} -9.938 
\times 10^{-5}i$\\
$C_{8}^{\prime} $   & $+3.839 \times 10^{-4}$                          &$ +3.839 \times 10^{-4} $\\
$C_{9}^{\prime} $   & $-1.023 \times 10^{-2} -2.963 \times 10^{-5}i$  &$-1.017 \times 10^{-2} -9.938 
\times 10^{-5}i$\\
$C_{10}^{\prime} $  & $+1.959 \times 10^{-3}$                          &$+1.959 \times 10^{-3}$\\
\hline
\hline
\end{tabular}
\end{center}
\caption{ Effective Wilson coefficients related to the tree operators, 
 electroweak and QCD penguin operators (see the reference  in text). }
\label{tab2}
\end{table}
%
%
\begin{table}
\begin{center}
\begin{tabular}{cccc} \hline \hline    
                          &      $\alpha$       &      $\beta$       &       $\gamma$   \\
\hline
\hline
 $(\rho_{min},\eta_{min})$     &       $ 104^{o}47$    &     $ 19^{o}32$    &       $  56^{o}21$    \\
\hline
 $(\rho_{min},\eta_{max})$  &       $ 93^{o}13$    &     $ 24^{o}31$ &      $ 62^{o}56$              \\
\hline  
 $(\rho_{max},\eta_{min})$  &       $ 112^{o}14$    &     $ 21^{o}20$ &      $ 46^{o}66$              \\       
\hline  
 $(\rho_{max},\eta_{max})$ &         $ 99^{o}66$    &     $ 26^{o}56$ &      $ 53^{o}78$              \\       
\hline  
\hline
\end{tabular}
\end{center}
\caption{Values of the CKM unitarity triangle  for limiting values of the CKM matrix elements.} 
\label{tab3}
\end{table}
%
%
\begin{table}[htp]
\begin{center}
\begin{tabular}{ccccccc} \hline \hline    
            &        &          & $B \rightarrow \rho$ &       &      &          \\
\hline
            &      $h_{V}$     & $h_{A_0}$ = $h_{A_3}$   &   $h_{A_1}$    &  $h_{A_2}$ & $m_{V}$ (${\rm GeV}^2$)& $m_{A_i}$ (${\rm GeV}^2$)  \\ 
\hline
 model $(1)$ & 0.329  & 0.281 & 0.283 & 0.283    & 5.32 & 5.32            \\
\hline
 model $(2)$ & 0.394 & 0.345 & 0.345 & 0.345   & 5.32 & 5.32         \\
\hline
\hline
            &        &          & $B \rightarrow \omega$ &       &      &          \\
\hline
            &      $h_{V}$     & $h_{A_0}$ = $h_{A_3}$   &   $h_{A_1}$    &  $h_{A_2}$ & $m_{V}$ (${\rm GeV}^2$)& $m_{A_i}$ (${\rm GeV}^2$)  \\ 
\hline
 model $(1)$ & 0.328  & 0.280 & 0.281 & 0.281    & 5.32 & 5.32            \\
\hline
 model $(2)$ & 0.394 & 0.345 & 0.345 & 0.345   & 5.32 & 5.32         \\
\hline
\hline
           &        &           & $B \rightarrow K^{*}$ &       &        &         \\
\hline
            &      $h_{V}$     & $h_{A_0}$ = $h_{A_3}$    &   $h_{A_1}$    &  $h_{A_2}$   & $m_{V}$ (${\rm GeV}^2$)& $m_{A_i}$ (${\rm GeV}^2$)\\ 
\hline
 model $(1)$ & 0.369 & 0.321 & 0.328 & 0.331       &  5.43  & 5.43              \\
\hline
 model $(2)$ & 0.443 & 0.360 & 0.402 & 0.416     &  5.43 & 5.43    \\
\hline
\hline
\end{tabular}
\end{center}
\caption{Form factor  values for  $B \rightarrow \rho$, $B \rightarrow \omega
  $ and  $ B \rightarrow K^{*}$ at $q^{2}=0$ (see the reference in text). }
\label{tab4}
\end{table}
%


    \begin{table} 
        \begin{center} 
            \begin{tabular}{cccc} 
               \hline 
                \hline
               channel & $\frac{q^2}{m_b^2}$ & BSW & GH \\ 
               \hline
                \hline 
                 &$0.3$ & $2.1$ & $1.0$\\
               $\bar{B}^0 \to \bar{K}^{* 0} \rho^0(\omega)$ &  & & \\ 
                 & $0.5$ & $1.5$ & $0.73$\\ 
               \hline  
                & $0.3$ & $6.6$ & $3.9$\\
               $B^- \to K^{* -} \rho^0(\omega)$ & & &\\ 
                & $0.5$ & $6.2$ & $3.6$\\ 
               \hline  
                \hline
                 & $0.3$ & $24$ & $13$\\
               $B^- \to \rho^{-} \rho^0(\omega)$ &  &  & \\ 
                & $0.5$ & $24$ & $14$\\ 
               \hline 
                \hline     
          \end{tabular} 
         \end{center}
\caption{$ \bar{B}^0, B^-$ branching ratios (in units of $10^{-6}$) using
                either the BSW or GH form factor models, for $q^2/m_b^2 = 0.3 (0.5)$, 
with $N_{c max}^{b \to s} = 2.84 (2.82)$, $N_{c max}^{b \to d} = 2.01 (1.95)$,
 $\rho = 0.229$ and $\eta = 0.325$.}
\label{tab5}
\end{table} 

        \begin{table}[p]
        \begin{center} 
            \begin{tabular}{cccc} 
               \hline 
                \hline
               channel & $\frac{q^2}{m_b^2}$ & BSW & GH \\ 
               \hline
                \hline 
                 &$0.3$ & $2.1$ & $1.0$\\
               $B^0 \to K^{* 0} \rho^0(\omega)$ &  & & \\ 
                 & $0.5$ & $1.7$ & $0.88$\\ 
               \hline  
                & $0.3$ & $5.8$ & $3.4$\\
               $B^+ \to K^{* +} \rho^0(\omega)$ & & &\\ 
                & $0.5$ & $3.8$ & $2.3$\\ 
               \hline  
                \hline
                 & $0.3$ & $20$ & $11$\\
               $B^+ \to \rho^{+} \rho^0(\omega)$ &  &  & \\ 
                & $0.5$ & $20$ & $11$\\ 
               \hline 
                \hline     
          \end{tabular} 
         \end{center}    
\caption{ $ B^0 , B^+$ branching ratios (in units of $10^{-6}$) using either the BSW or GH form factor models, for $q^2/m_b^2 = 0.3 (0.5)$, 
with $N_{c max}^{b \to s} = 2.84 (2.82)$, $N_{c max}^{b \to d} = 2.01 (1.95)$,
 $\rho = 0.229$ and $\eta = 0.325$.}
\label{tab6}     
\end{table} 
    \begin{table}[p]
        \begin{center} 
            \begin{tabular}{cccc} 
               \hline 
                \hline
               channel & $\frac{q^2}{m_b^2}$ & BSW & GH \\ 
               \hline
                \hline 
                 &$0.3$ & $+0.36$ & $-0.45$\\
                 $  \bar{K}^{* 0}(K^{* 0}) \rho^0(\omega)$ &  & & \\
                 & $0.5$ & $+4.70$ & $+5.90$\\ 
               \hline  
                & $0.3$ & $-6.6$ & $-6.37$\\
               $ K^{* -}(K^{* +})  \rho^0(\omega)$ & & &\\
                & $0.5$ & $-23.0$ & $-22.0$\\ 
               \hline 
                \hline 
                 & $0.3$ & $-8.5$ & $-9.6$\\
               $ \rho^{-}(\rho^{+})  \rho^0(\omega)$ &  &  & \\ 
                & $0.5$ & $-8.7$ & $-9.9$\\ 
               \hline 
                \hline     
          \end{tabular} 
         \end{center} 
         \caption {Global $CP$-violating asymmetries (in percents) 
                  using either the BSW or GH form factor models, for $q^2/m_b^2 = 0.3 (0.5)$, 
         with $N_{c max}^{b \to s} = 2.84 (2.82)$, $N_{c max}^{b \to d} = 2.01 (1.95)$, $\rho = 0.229$ and
         $\eta = 0.325$.}
         \label{tab7}  
      \end{table}



%
\end{document}